\documentclass[12pt,preprint]{aastex}
\usepackage{graphicx}
\usepackage{subfigure}
\usepackage {natbib}

\begin{document}

\title{Spitzer Observations of White Dwarfs: the Missing Planetary Debris Around DZ Stars}

\author{S. Xu(\includegraphics[width=1.2cm]{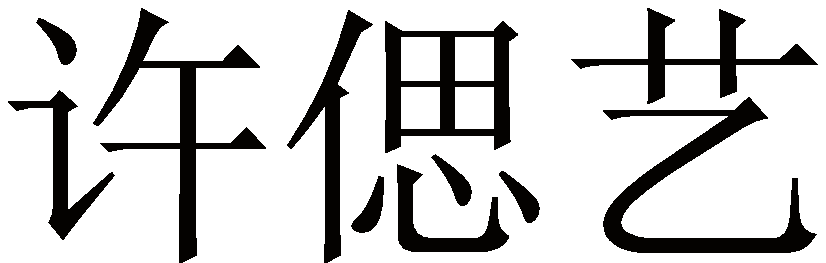})\altaffilmark{a}, M. Jura\altaffilmark{a}}
\altaffiltext{a}{Department of Physics and Astronomy, University of California, Los Angeles CA 90095-1562; sxu, jura@astro.ucla.edu}

\begin{abstract}
We report a Spitzer/IRAC search for infrared excesses around white dwarfs, including 14 newly-observed targets and 16 unpublished archived stars. We find a substantial infrared excess around two warm white dwarfs \--- J220934.84+122336.5 and WD 0843+516, the latter apparently being the hottest white dwarf known to display a close-in dust disk. Extending previous studies, we find that the fraction of white dwarfs with dust disks increases as the star's temperature increases; for stars cooler than 10,000 K, even the most heavily polluted ones do not have $\sim$1000 K dust. There is tentative evidence that the dust disk occurrence is correlated with the volatility of the accreted material. In the Appendix, we modify a previous analysis to show that Poynting-Robertson drag might play an important role in transferring materials from a dust disk into a white dwarf's atmosphere.

\end{abstract}


\section{INTRODUCTION}

White dwarfs are the final evolutionary stage of stars with masses less than about eight solar masses. The first infrared excess around a white dwarf was discovered more than 20 years ago \citep{ZuckermanBecklin1987}. Since 2005, primarily thanks to the {\it Spitzer Space Telescope} \citep{Werner2004}, progress has been dramatic. Including WD 0843+516\footnote{After reading a short description on astro-ph of a survey on dust disks around white dwarfs \citep{Chu2010}, we learned from Y.-H. Chu that their team has independently found the infrared excess for this star.} and J220934.84+122336.5 (hereafter J2209+1223), which are reported in this paper, there are 23 confirmed white dwarfs with a warm disk and 5 additional candidates (see our Table \ref{Tab: NewWD}, an extension of Table 1 in \citet{Farihi2009}). Recently, observations from the WISE mission \citep{Debes2011} and the UKIRT Infrared Deep Sky Survey (UKISS) \citep{Steele2011, Girven2011} are also contributing to the disk studies. 

There are three classes of dust disks detected around a white dwarf: (1) $\sim$100 K dust, which is found around $\gtrsim$ 15\% pre-white dwarfs and white dwarfs, sometimes in the center of a planetary nebulae \citep{Chu2011}. The source of the dust is believed to be the collisions among analogs to Kuiper-belt objects in the system \citep{BonsorWyatt2010, Dong2010}. (2) $\sim$500 K dust around two cool stars, G 166-58 (T$_*$=7,400 K) \citep{Farihi2008b} and PG 1225-079 (T$_*$=10,600 K) \citep{Farihi2010b}. A model that reproduces these data invokes emission from an opaque dust ring with a large, dust-free inner hole and an outer boundary within the tidal radius of the white dwarf. However, there are only two dust rings in this class and their origin is uncertain. (3) $\sim$1000 K dust, which is found in the remaining 21 white dwarfs listed in Table \ref{Tab: NewWD}. In these systems, all the dust lies within the star's tidal radius and the inner boundary is determined by the location where the refractories rapidly sublimate. The excess can be fit with a flat opaque dust disk \citep{Jura2003}. The excesses from 1000 K dust are substantial at wavelengths greater than 3 $\mu$m and can be easily studied with warm {\it Spitzer}; we focus on discussing this class of disk in this paper\footnote{In the following sections, unless specifically stated, dust disk and infrared excess all refer to this 1000 K dust.}.

Gravitational settling of heavy elements is usually so effective in white dwarfs cooler than 25,000 K that their atmospheres are typically pure hydrogen or pure helium \citep{Koester2009a}. However, about 25\% of DAs (white dwarfs with hydrogen dominated atmosphere) and 33\% of DBs (white dwarfs with helium dominated atmosphere) have atmospheric pollution \citep{Zuckerman2003, Zuckerman2010}. Given that the diffusion timescale of heavy elements is orders of magnitude less than a white dwarf's cooling age \citep{Koester2009a}, it is unlikely that those heavy elements are intrinsic. Theoretical calculations show that tidal disruption and subsequent accretion of small bodies in the planetary system can explain the source of pollution in these objects \citep{DebesSigurdsson2002, Jura2003, Bonsor2011}. \citet{Kilic2006} first showed a correlation between the presence of an orbiting dust disk and a high level of heavy element pollution in the star's photosphere. Subsequent observations have confirmed this trend and supported the small body accretion model \citep{Jura2007b, Farihi2008a, Farihi2008b, Farihi2009, Farihi2010b}. Thus studying polluted white dwarfs provides invaluable measures of the bulk composition of extrasolar minor planets (see, for example, \citet{Klein2010}).

However, as can be seen in our Table \ref{Tab: NewWD} and in \citet{Farihi2009}, almost all the disk-host stars have effective temperatures of at least 9,500 K. Here, we report the results from a warm {\it Spitzer} Cycle 7 program, searching for infrared excesses around polluted cool white dwarfs and a few highly-polluted warmer targets. Additionally, we reduced some unpublished archived {\it Spitzer} data.

Target selections and observations are presented in section 2. In section 3, we describe data reduction procedures and SED fits to the data. In section 4, the main result is discussed: we continue to find dust disks around warm polluted white dwarfs but no dust around highly polluted cool stars. In section 5, we explore the possible correlations between the presence of a dust disk and other characteristics of the star in addition to its temperature. Possible scenarios that can account for the presence of heavy elements without producing a dust disk are discussed in section 6. Conclusions are presented in section 7. In the Appendix, we assess the importance of Poynting-Robertson (P-R) drag on the entire dust disk as a mechanism to transfer material from the disk into the white dwarf's atmosphere.

\section{OBSERVATIONS}

\subsection{Source Selection}

The 14 targets of our Cycle 7 program are listed in Table \ref{Tab: Target}. 10 stars are newly identified DZs\footnote{DZs are cool white dwarfs that display trace elements other than carbon; it is thought that DZs have a helium dominated atmosphere, but the gas is too cool for the helium lines to be detected.} with distinctively large amount of atmospheric calcium detected in the SDSS spectrum; they all have T$_*$ $<$ 10,000 K and m(z) $<$ 18.0 mag (0.23 mJy) \citep{Dufour2006, Dufour2007}. 4 additional DBZs\footnote{DBZs are DBs with detected heavy elements in the atmosphere.} with T$_*$ $\sim$ 15,000 K are also included: 3 stars from \citet{Eisenstein2006} plus G 241-6, which was recognized recently to be a near-twin of GD 40, a highly polluted white dwarf with a dust disk \citep{Zuckerman2010, Klein2010}. 

For the completeness of the sample, we also report IRAC 3.6 $\mu$m -- 7.9 $\mu$m fluxes for 16 single DA white dwarfs from the unpublished archived Cycle 3 program 30856. These stars all have well determined temperatures, surface gravities as well as highly accurate SDSS photometry, which is crucial in determining the presence of a disk. Relevant parameters are listed in Table \ref{Tab: ArchiveTarget}.

\subsection{Observing Strategies}

{\it Spitzer} has been operated in the warm phase mission since the depletion of cryogen in May, 2009. The two shortest wavelength channels (3.6 $\mu$m and 4.5 $\mu$m) of the IRAC \citep{Fazio2004} continue to function and were used in this study to obtain broadband images. The observations were performed by using a 30 sec frame time with 30 medium size dithers in the cycling pattern, resulting in a 900 sec total exposure time in each IRAC channel.

The archived data were obtained between 2006 and 2007. The observing mode was random 9-point large dithering with a frame time of 30 sec, resulting in a total of 270 sec in each IRAC channel.

\section{DATA ANALYSIS}

\subsection{Data Reduction and SED Fits}

Following the data reduction procedures described in \citet{Farihi2008a, Farihi2008b,Farihi2009}, each exposure was processed with MOPEX (version 18.4.9) to create a single mosaic with a pixel size of 0{\farcs}6. Though the intrinsic plate scale is 1{\farcs}2 pixel$^{-1}$, a 0{\farcs}6 pixel$^{-1}$ mosaic was used because it has a better spatial resolution and the 3.6 $\mu$m and 4.5 $\mu$m images are oversampled. Aperture photometry was performed on the combined mosaic using both: (i) the standard IRAF apphot task, and (ii) the Astronomical Point source EXtractor (APEX) in MOPEX. We tried an aperture size of both 2 (2{\farcs}4) and 3 (3{\farcs}6) native pixels with a sky annulus of an inner radius of 5 native pixels (6{\farcs}0) and an outer radius of 15 native pixels (18{\farcs}0). We found that in some cases even in a clean field, the measured flux in these two aperture sizes can differ up to 5\% and we report the average flux weighted by their SNR. To derive the aperture correction factor, we performed aperture photometry in the point response function (PRF) images in each band with our set of parameters. The PRF is the convolution of the point response function (PSF) of the telescope and the pixel response function of the detector and it is provided by the Spitzer Science Center. There are three factors contributing to the total uncertainty: 5\% calibration uncertainty, which is based on previous studies of white dwarfs \citep{Farihi2008a}, rather than the 2-3\% value optimally derived for other targets \citep{Reach2005b, Bohlin2011}, measurement error and uncertainties caused by the choice of aperture radius; these uncertantities were all added in quadrature. The fluxes independently obtained by IRAF and APEX agree to within 3\%. To be conservative, we report the values that have the larger measurement error in Table \ref{Tab: Flux} and Table \ref{Tab: ArchiveFlux}.

When there is a background source, proper motion analysis is performed to accurately determine the position of the target, followed by PRF fitting to obtain the true flux of the star. A residual mosaic, which subtracted the detected point source from the original mosaic, was produced. To check the result of PRF fitting, aperture photometry was re-performed at the same location of the target in the residual image. A good fit is achieved if the flux level in the aperture is comparable to the sky level, which is a few percent of the flux of the source. Examples of PRF fitting and the residual image are shown in Figure \ref{Fig: PSF} and Figure \ref{Fig: psf_G241-6}.

For some archived stars, the background is very noisy in the 5.7 $\mu$m and 7.9 $\mu$m band. We report a positive detection when the measured flux is at least 3$\sigma$. To measure the upper limit, aperture photometry was performed in 20 empty background regions around the source with an aperture size of 2 native pixels. Then the standard deviation of these fluxes is taken to equal 1$\sigma$.

\begin{figure}
\epsscale{1.2}
\plottwo{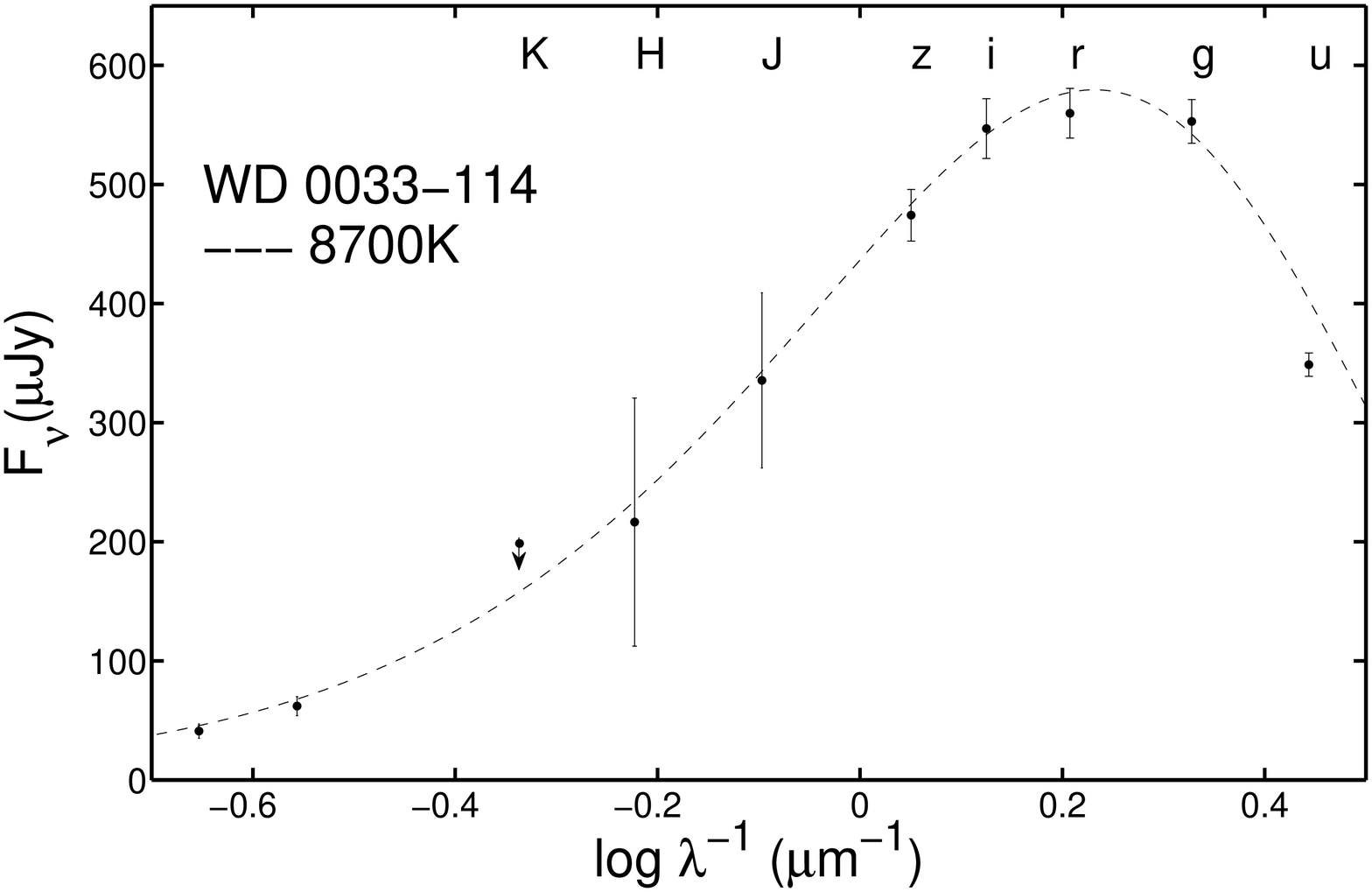}{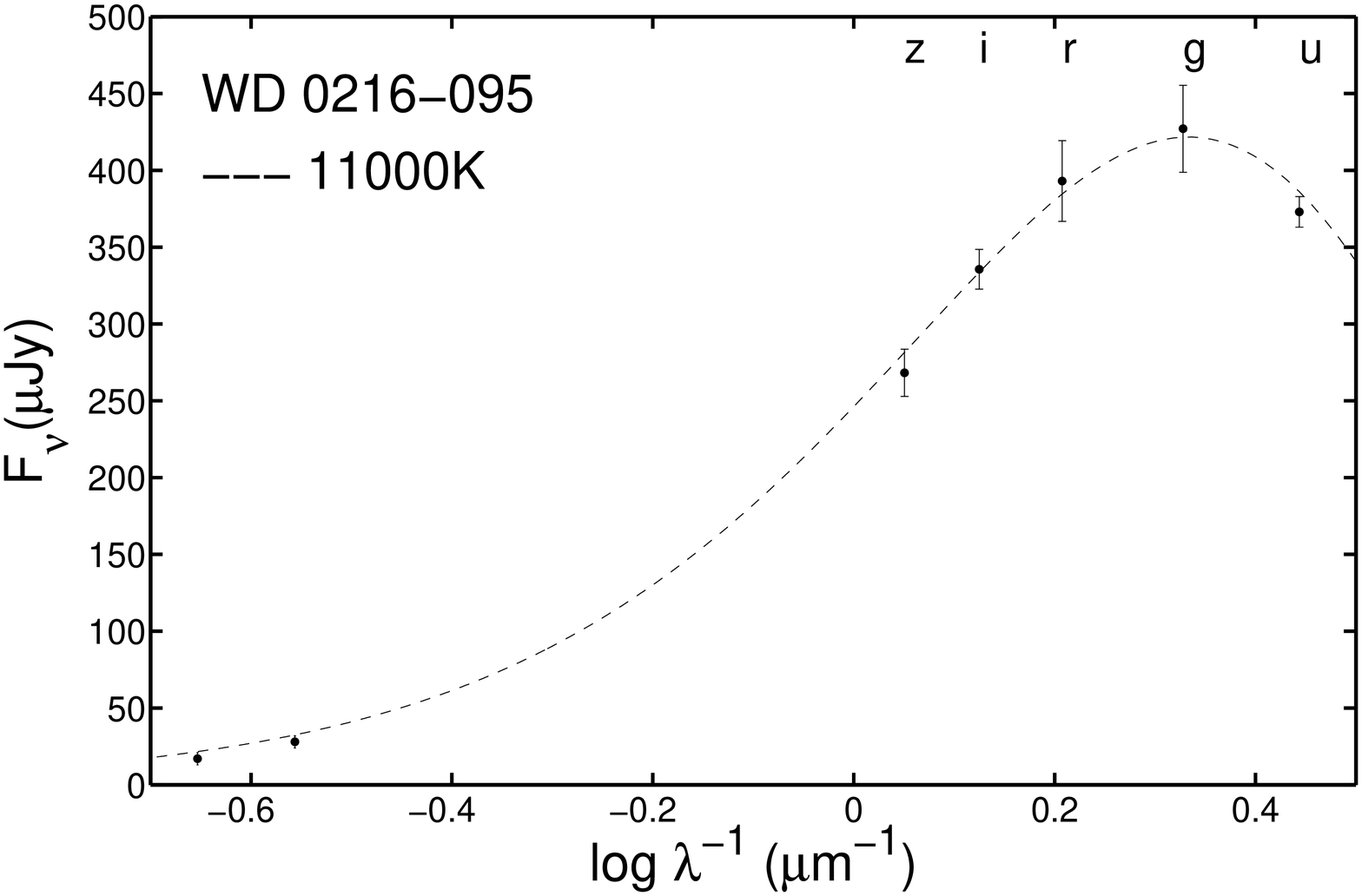}
\plottwo{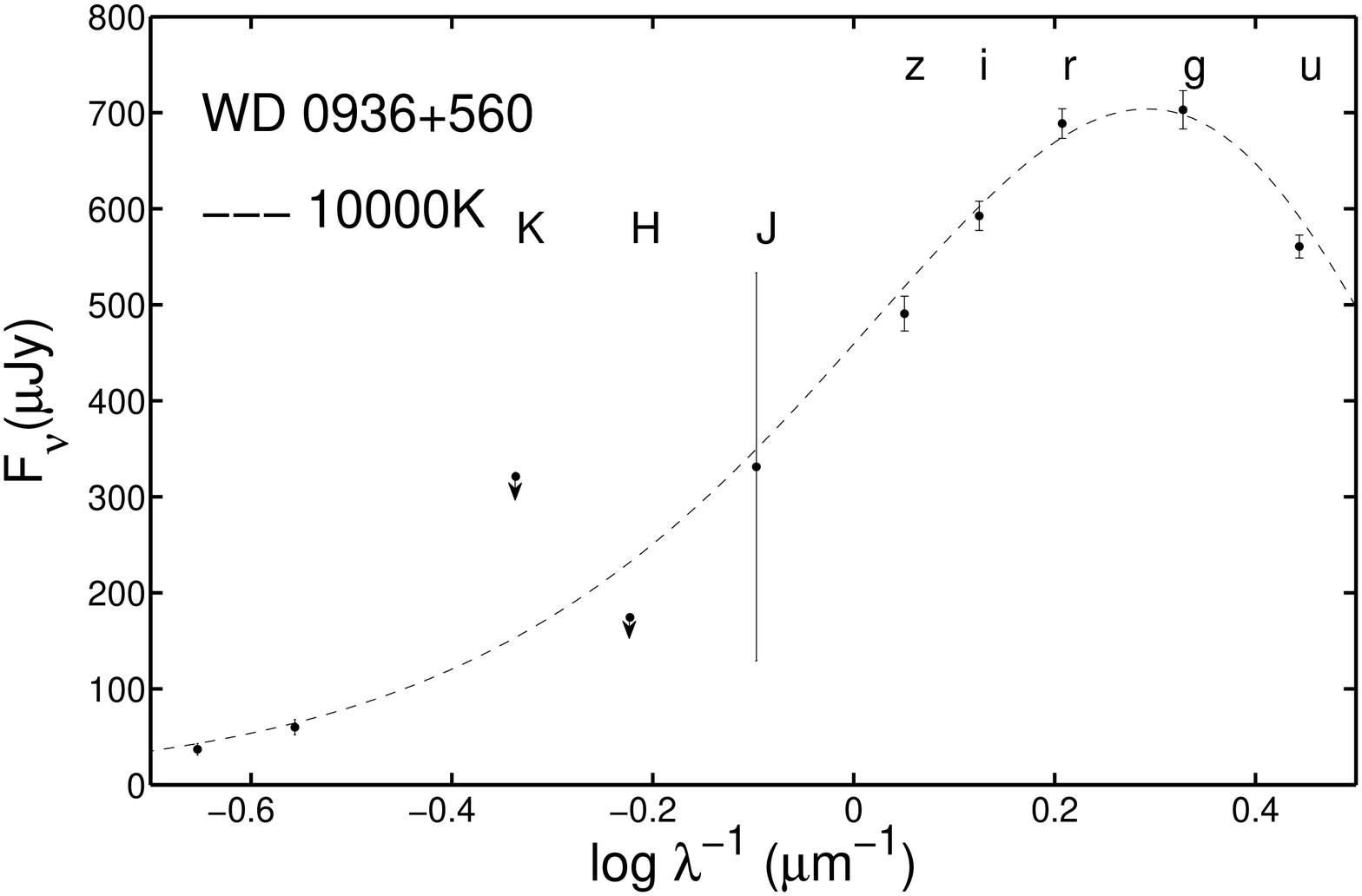}{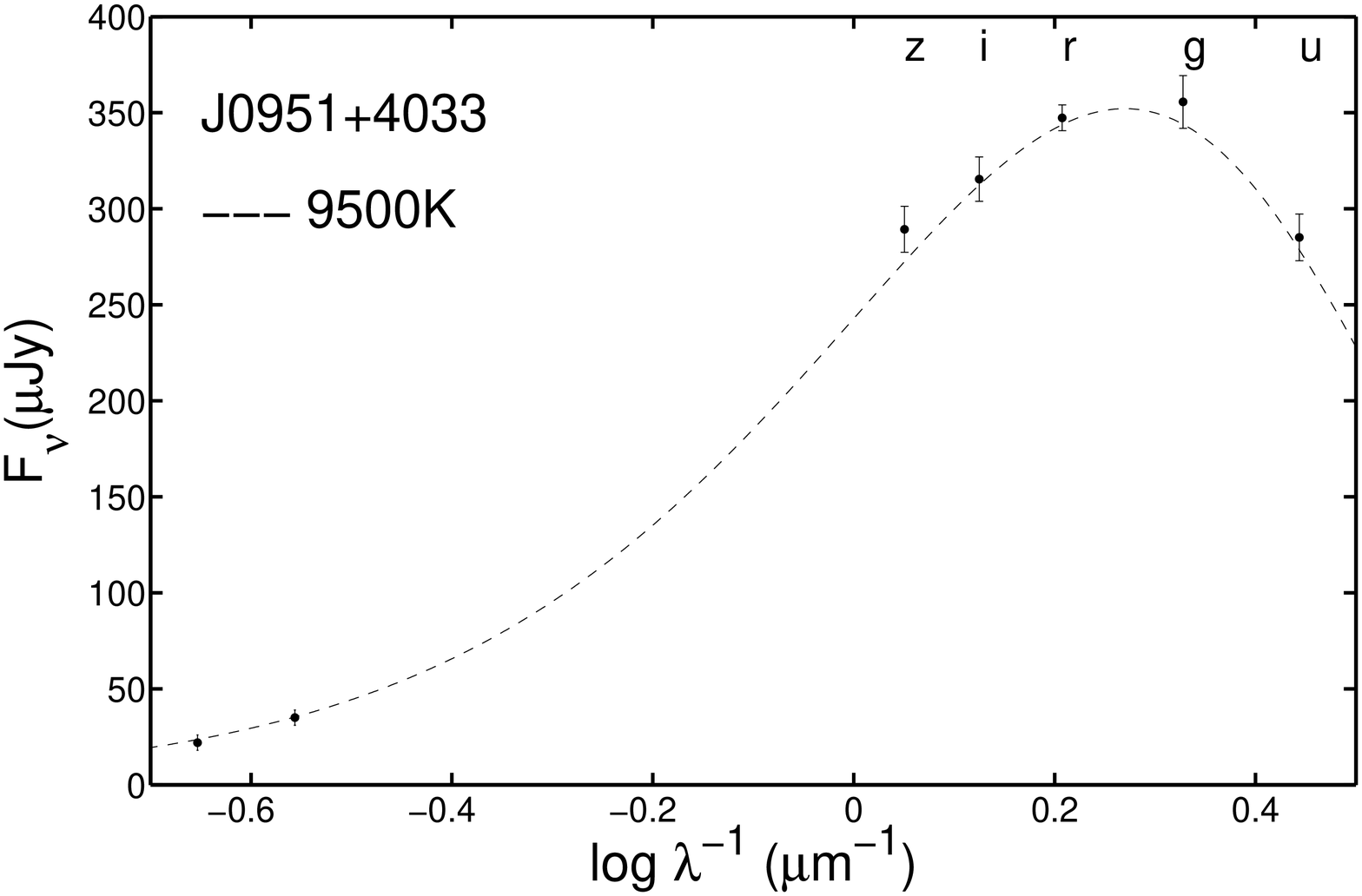}
\plottwo{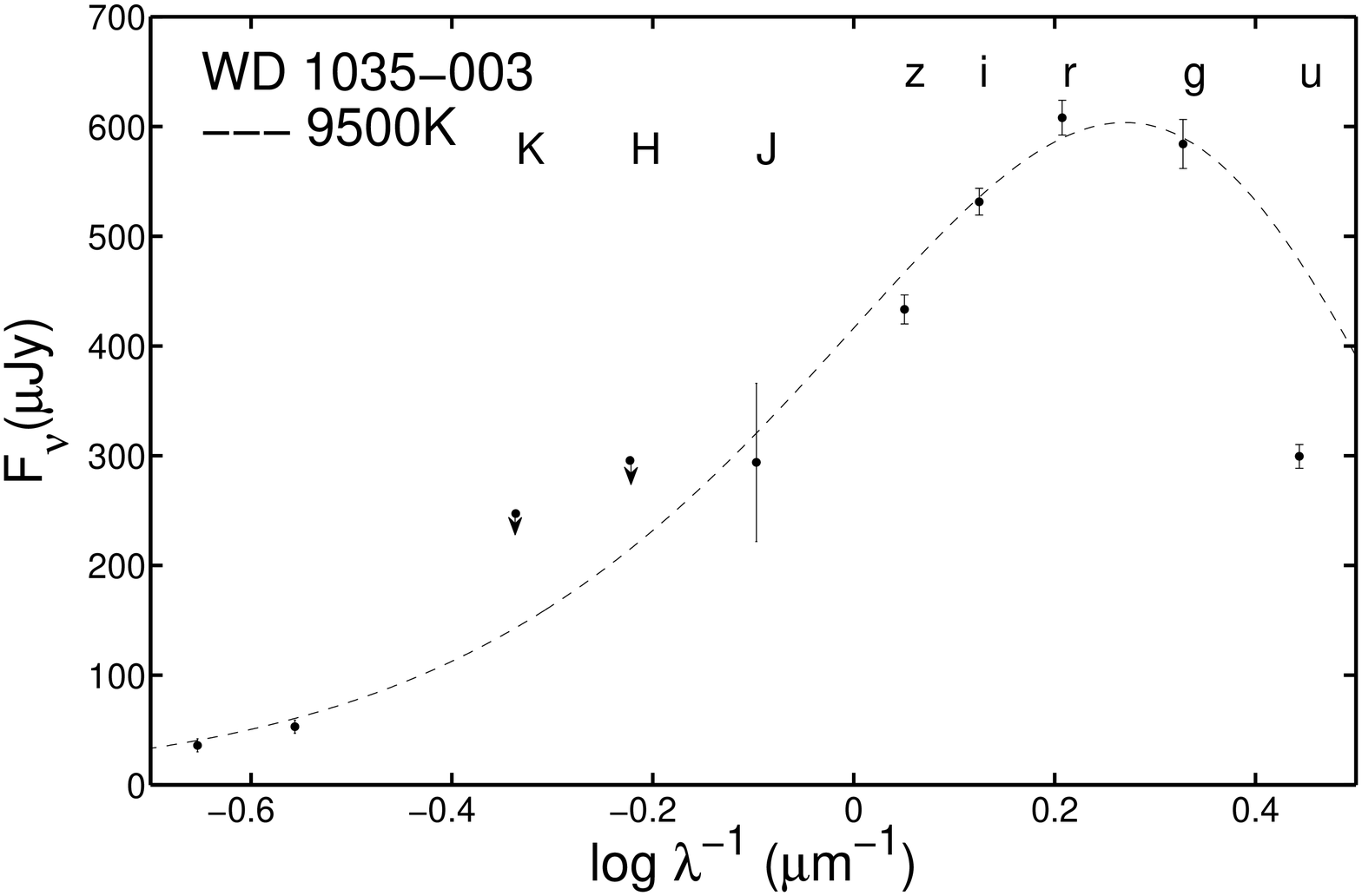}{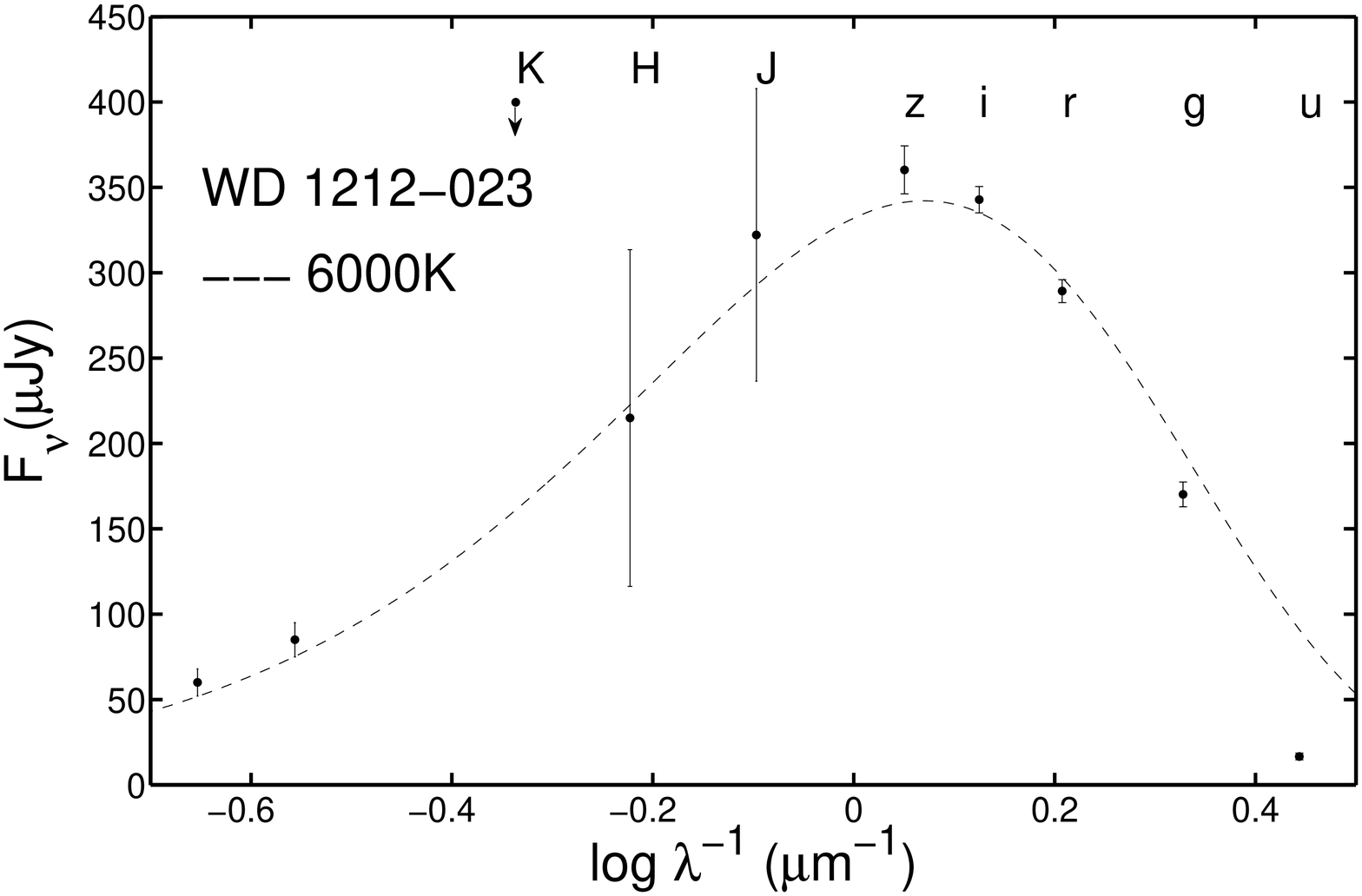}
\caption{SED for Cycle 7 targets, including data from SDSS, 2MASS and IRAC with 2$\sigma$ error bars. The dashed line is a simple blackbody fit to the photospheric flux.}\label{Fig: SED7a}
\end{figure}

\begin{figure}
\epsscale{1.2}
\plottwo{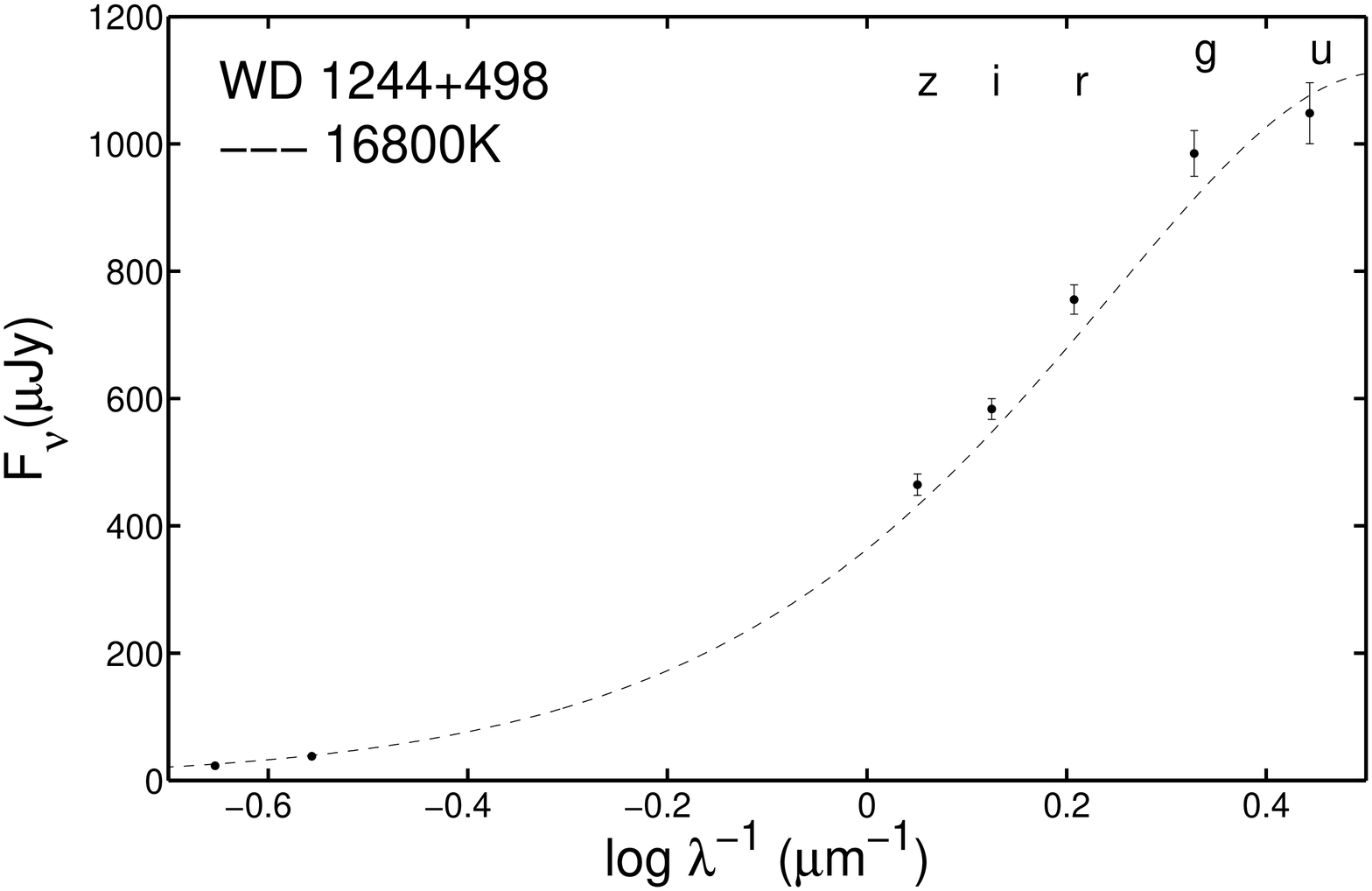}{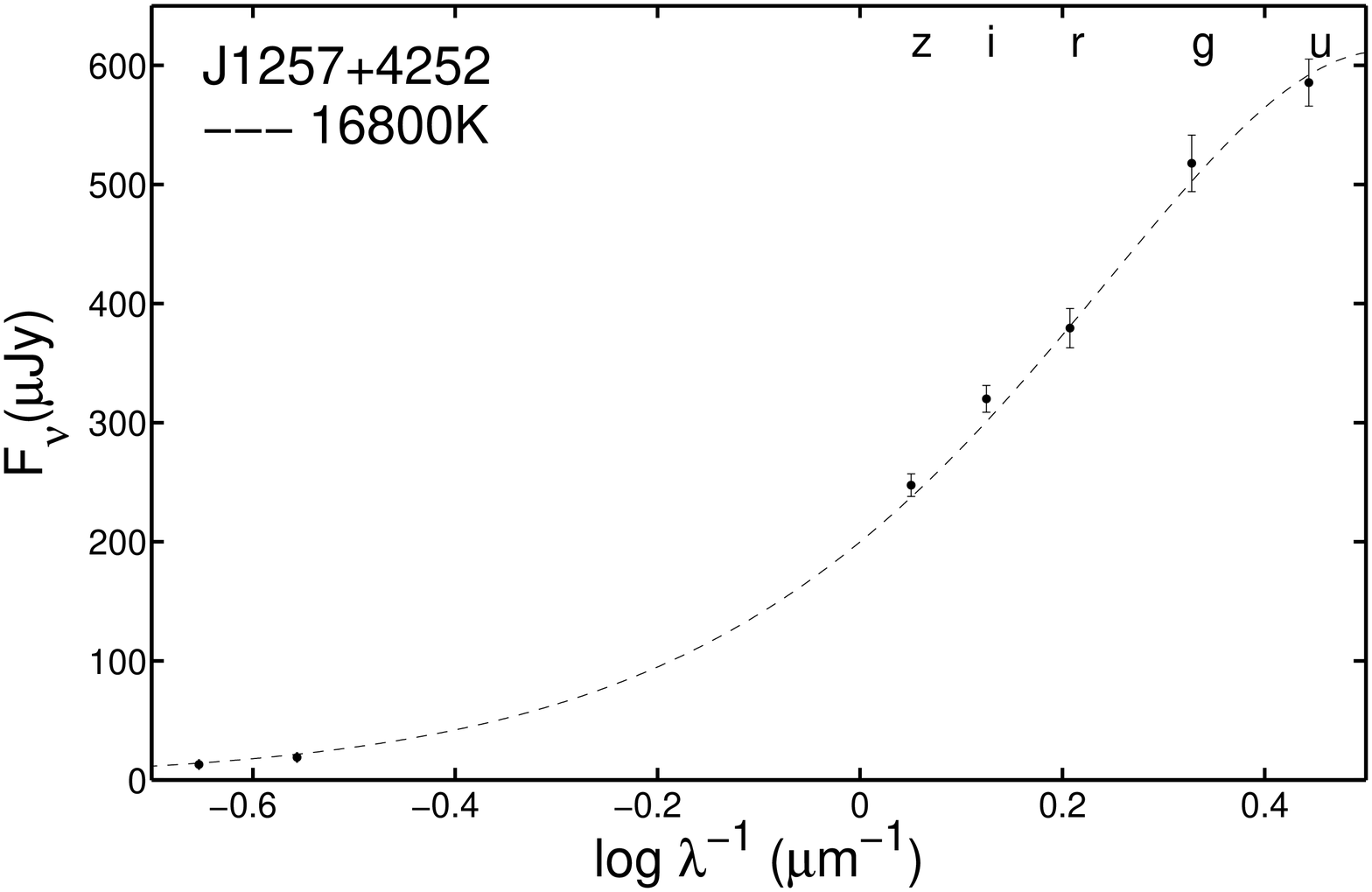}
\plottwo{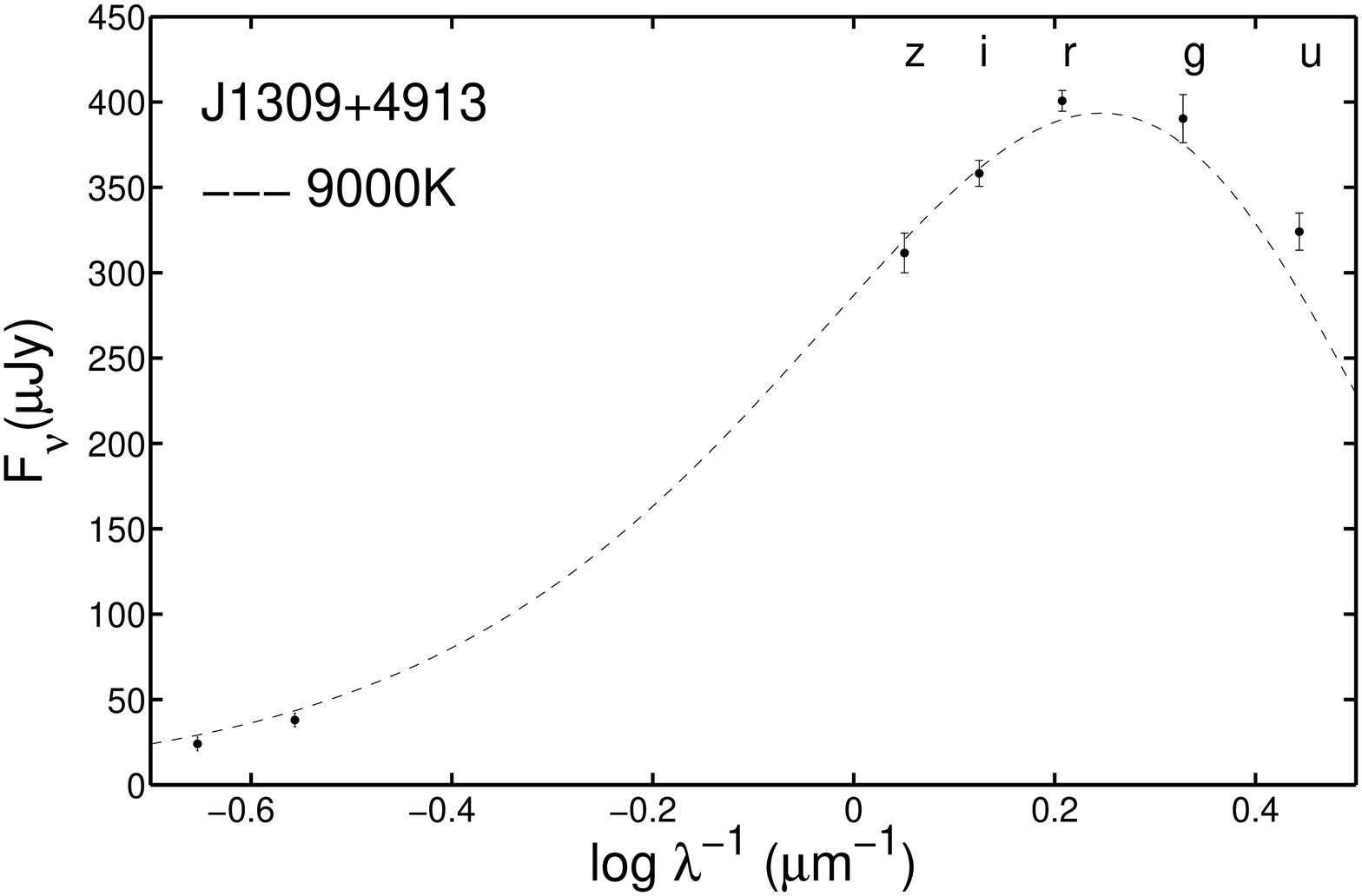}{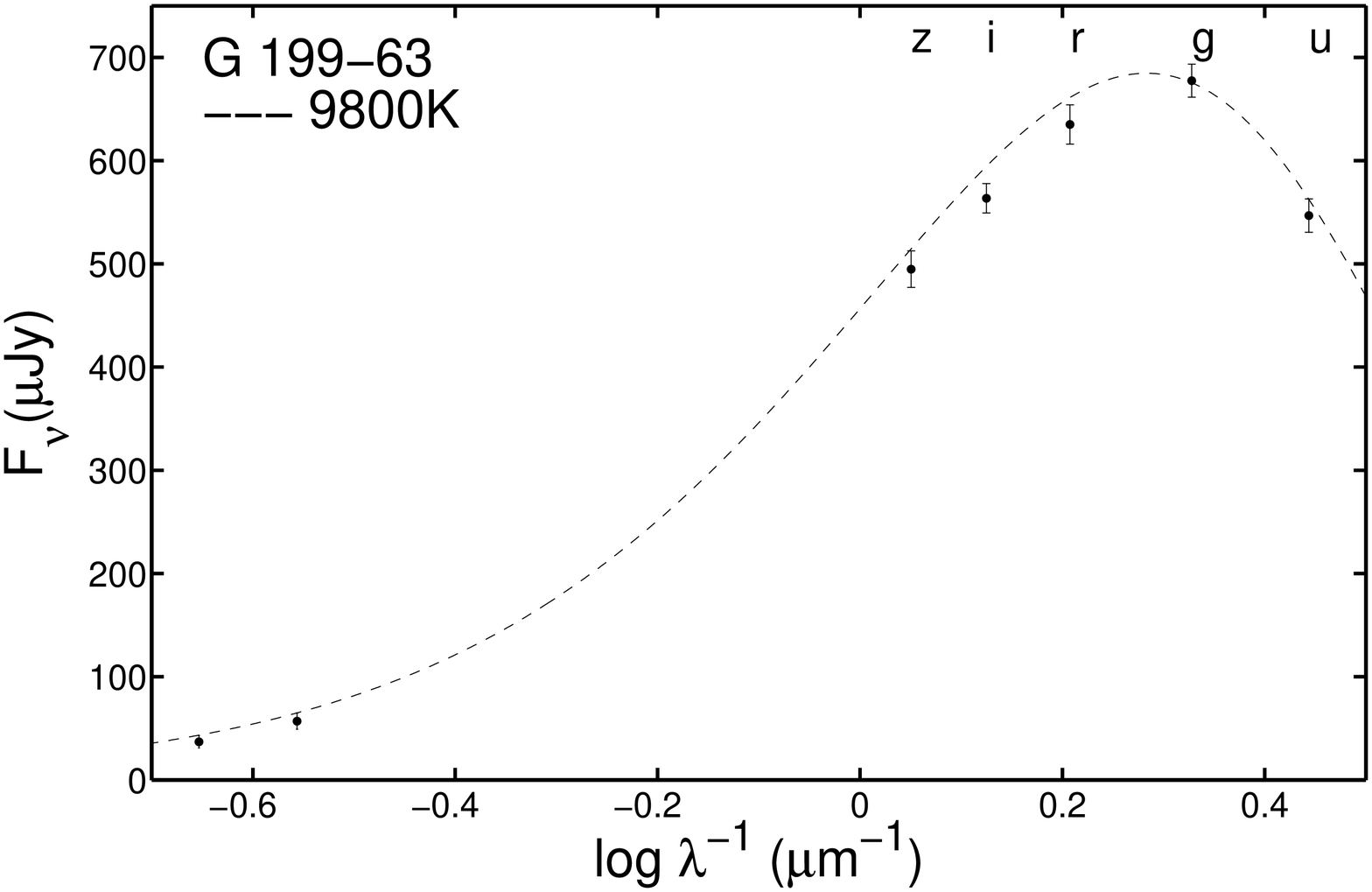}
\plottwo{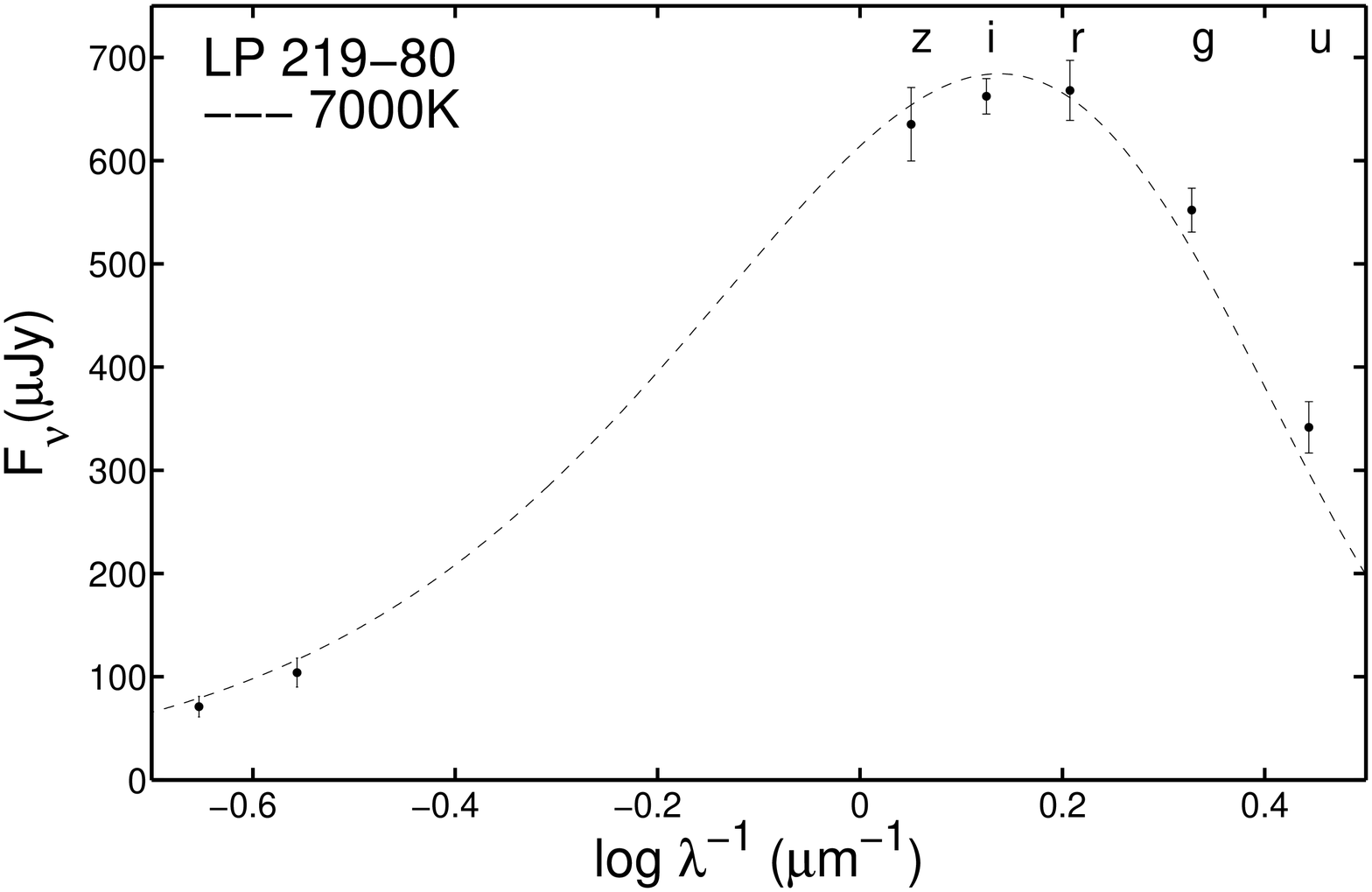}{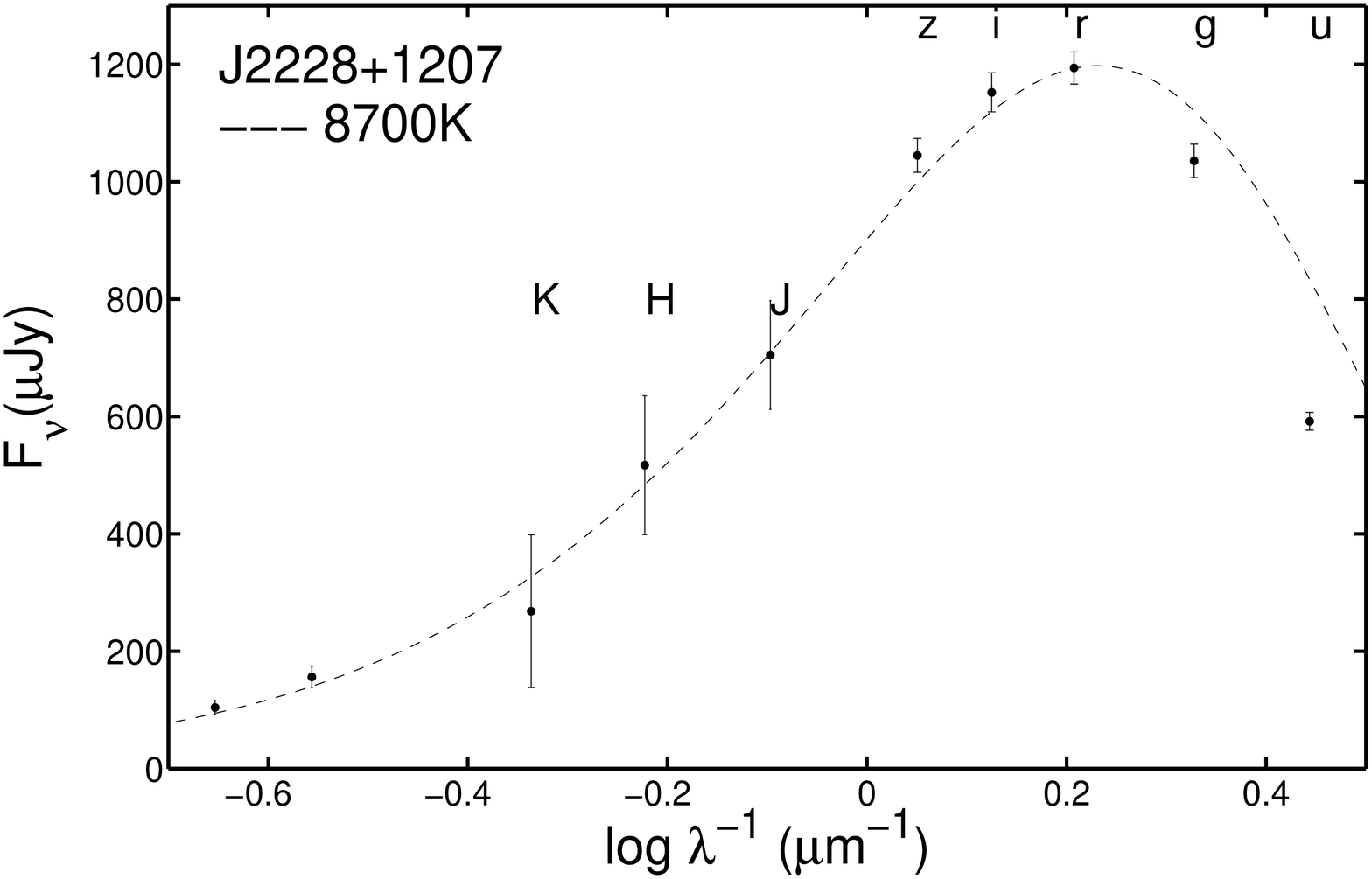}
\caption{The same as Figure \ref{Fig: SED7a}.}\label{Fig: SED7b}
\end{figure}

\begin{figure}[tbph]
\epsscale{0.8}
\plotone{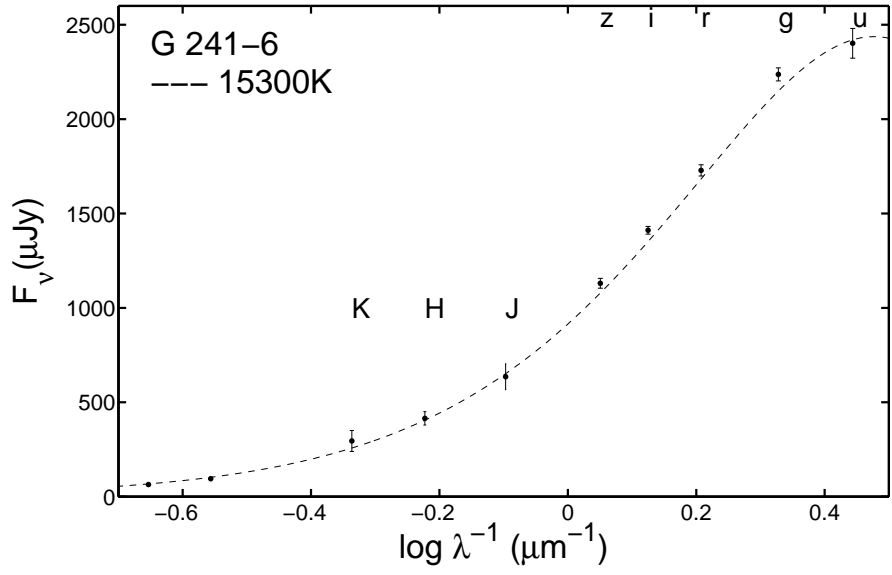}
\caption{The same as Figure \ref{Fig: SED7a} except for G 241-6.  \label{Fig: G241-6}}
\end{figure}

\begin{figure}
\epsscale{1.2}
\plottwo{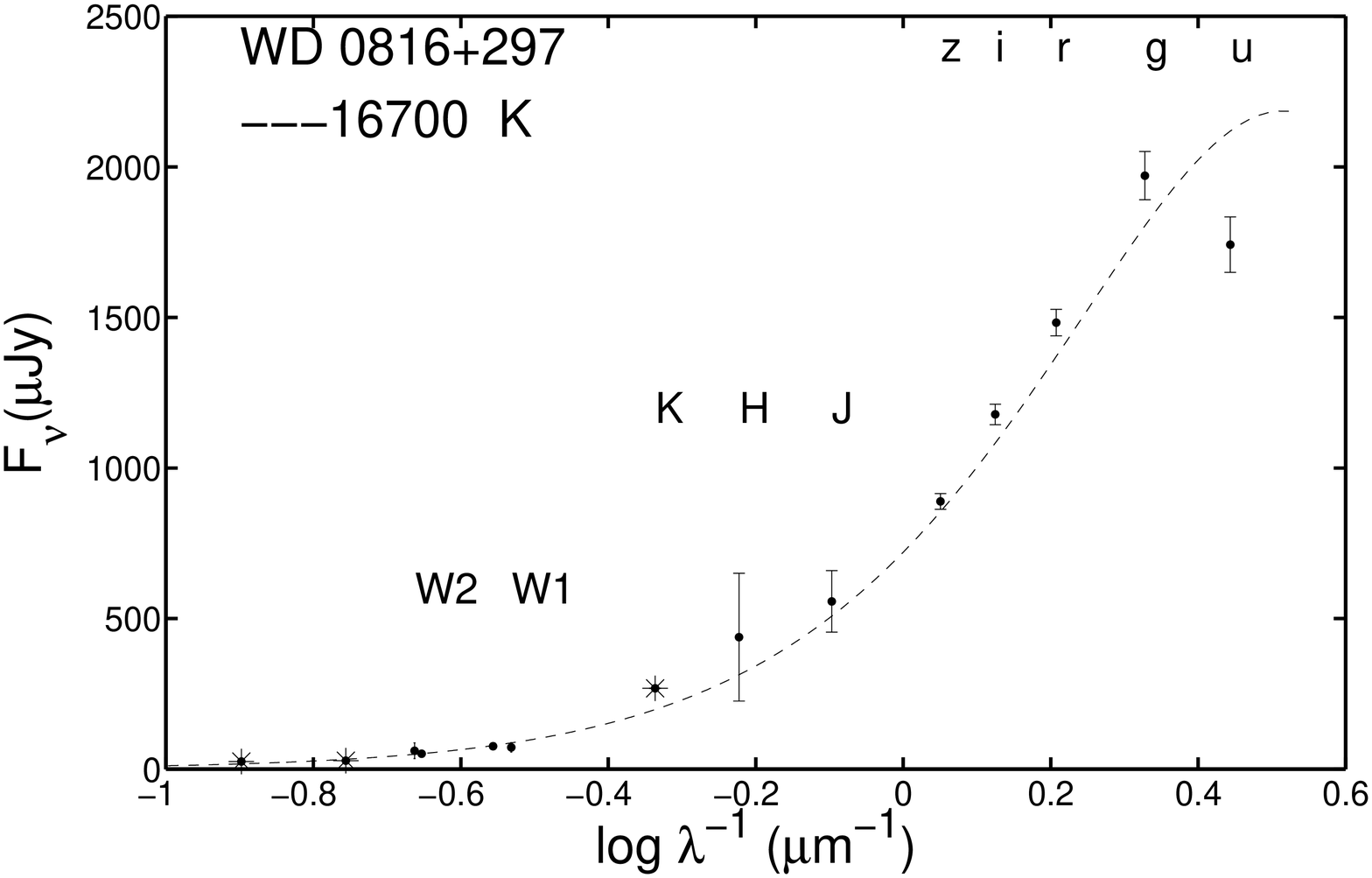}{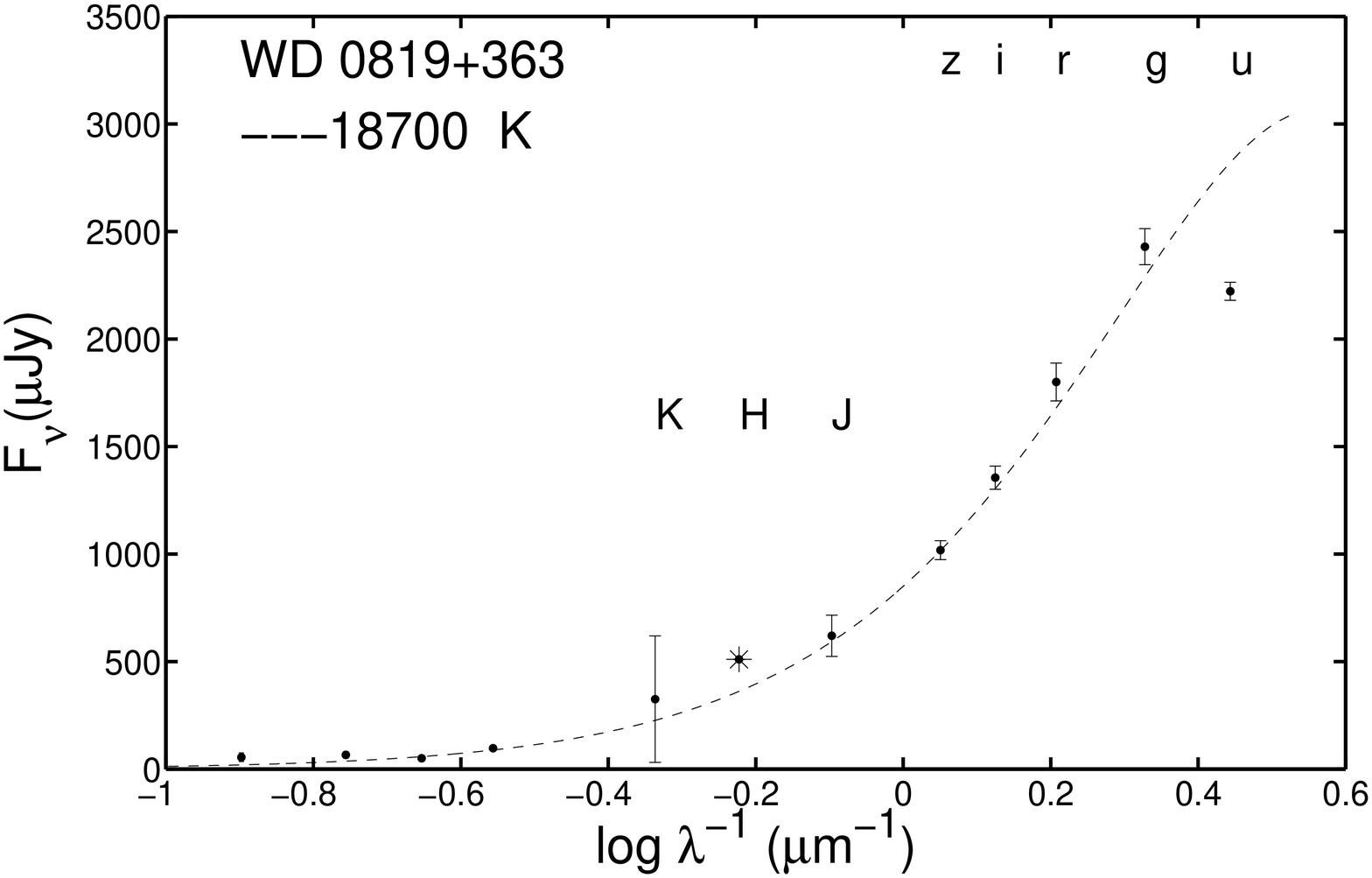}
\plottwo{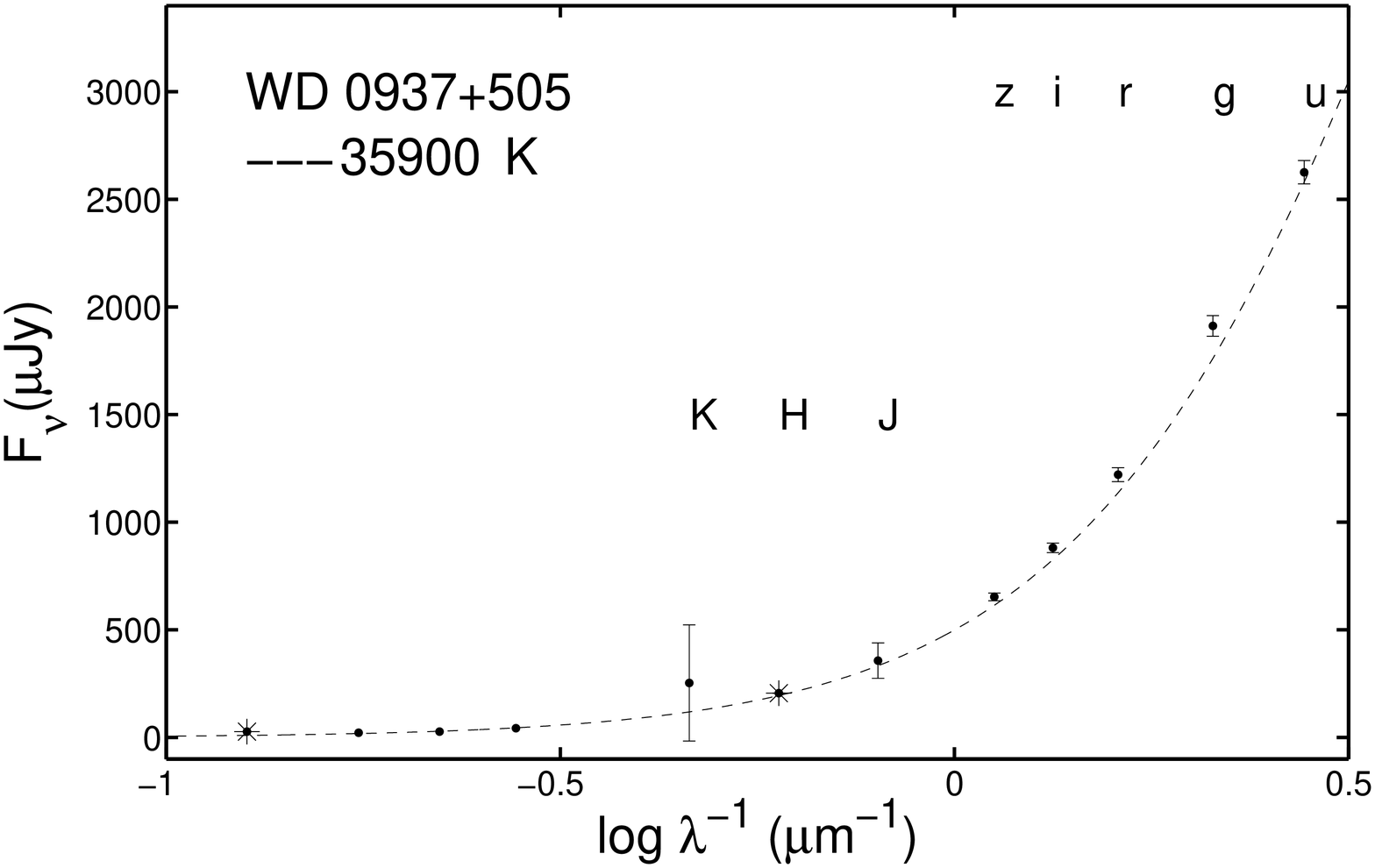}{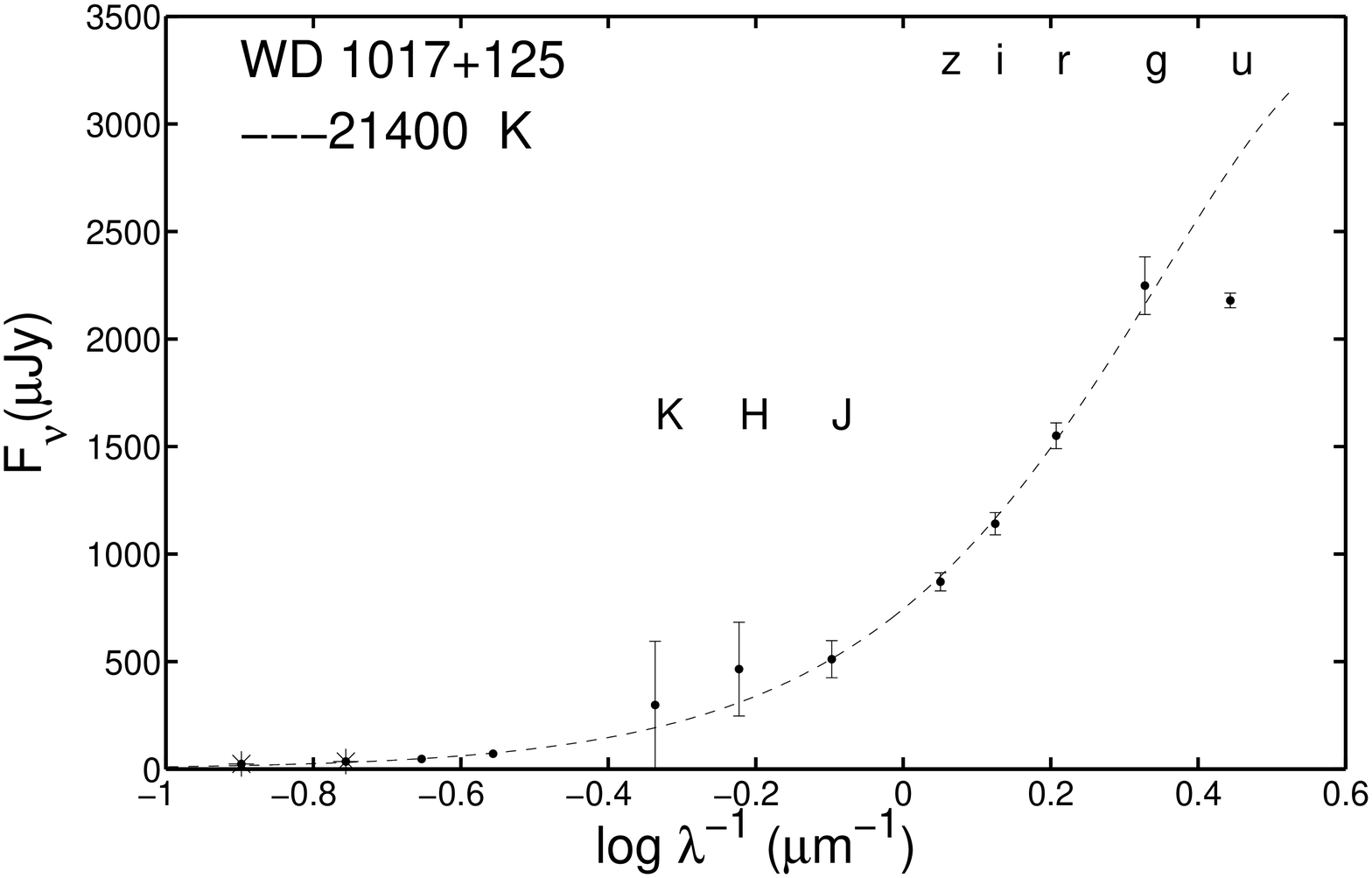}
\plottwo{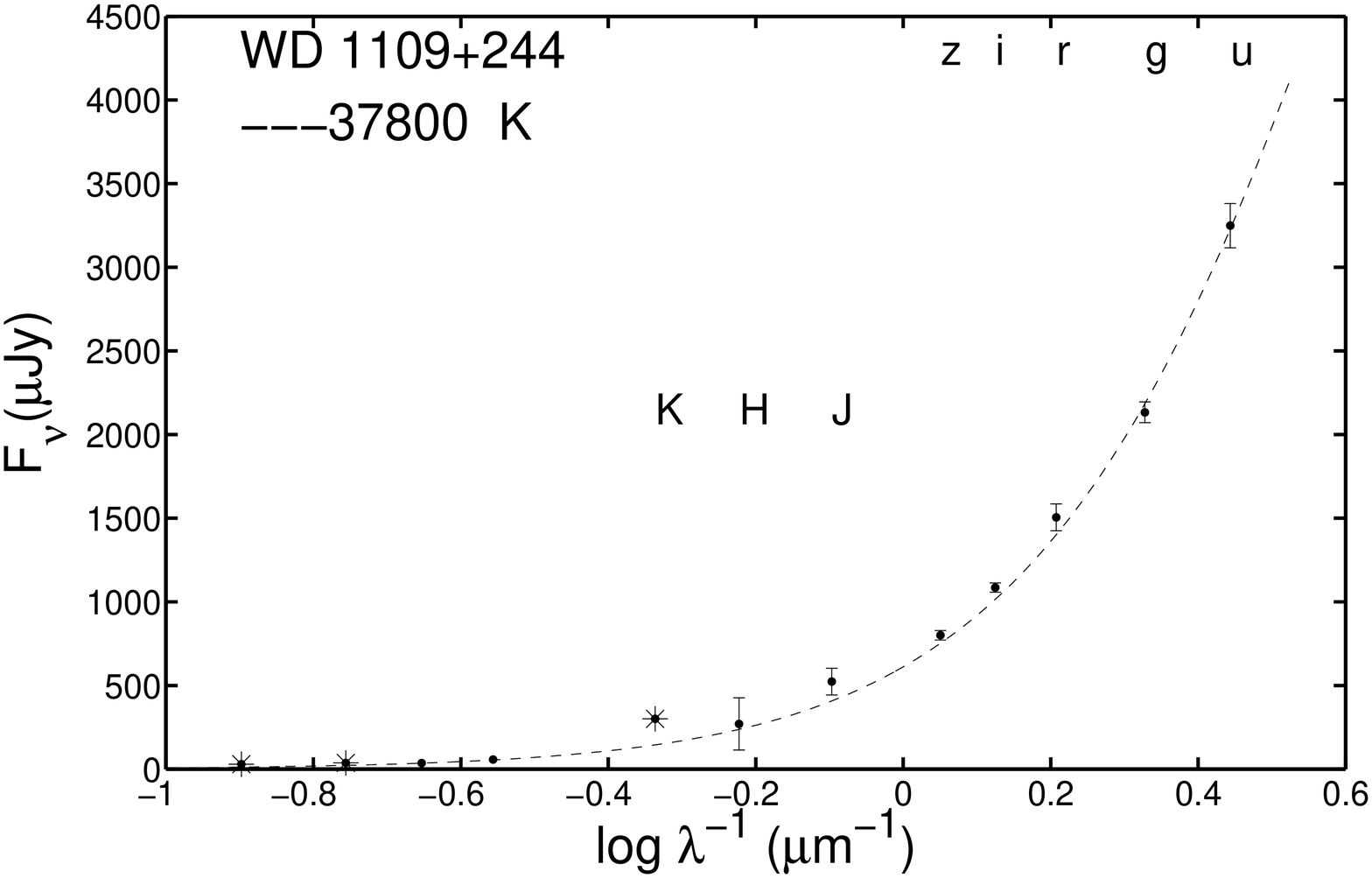}{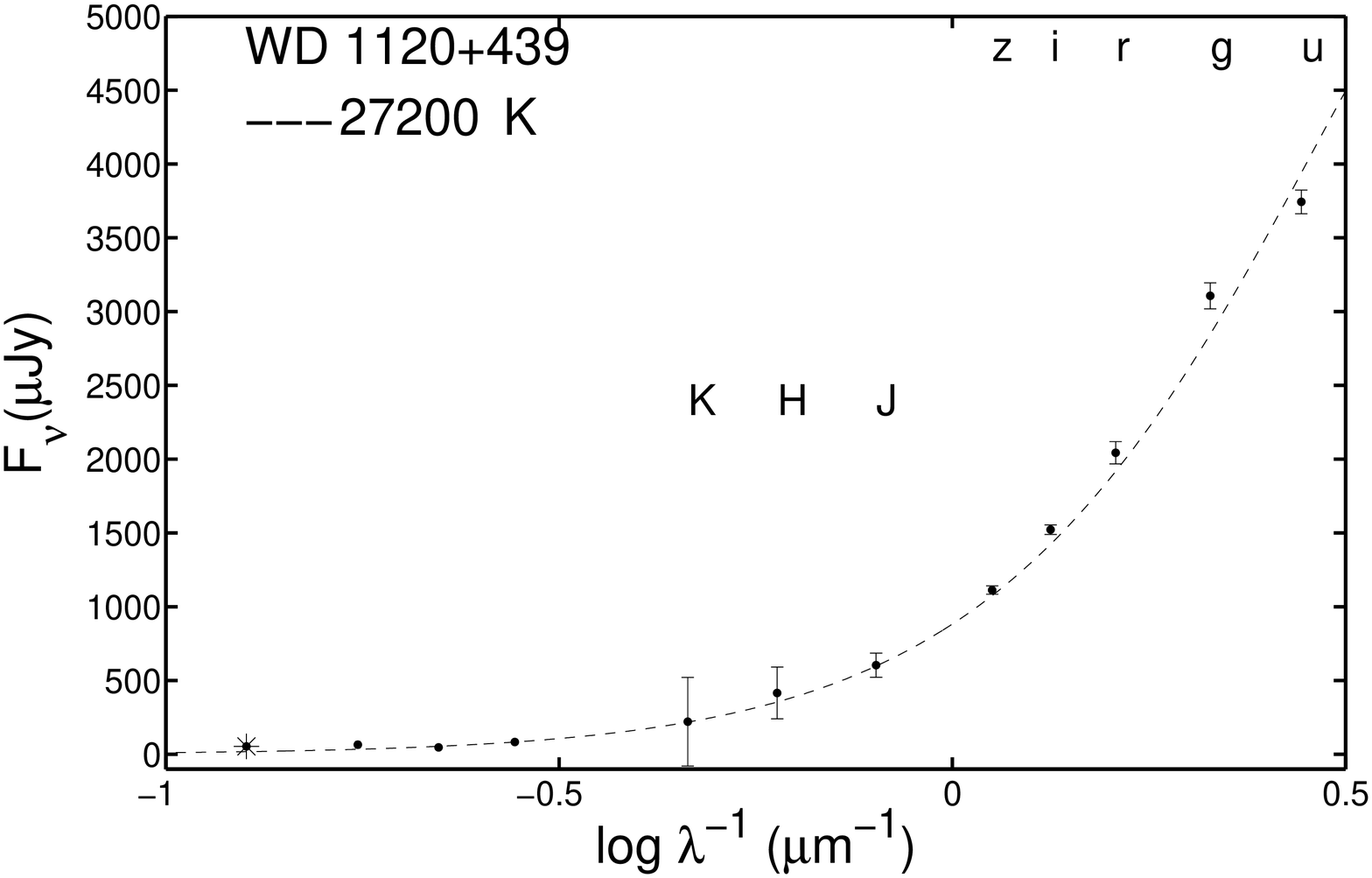}
\caption{SED for archived targets, including data from SDSS, 2MASS, IRAC and WISE data when available with 2$\sigma$ error bars. The asterisks denote upper limit. The dashed line is a simple blackbody fit to the photospheric flux.} \label{Fig: ArchiveSED1}
\end{figure}

\begin{figure}
\epsscale{1.2}
\plottwo{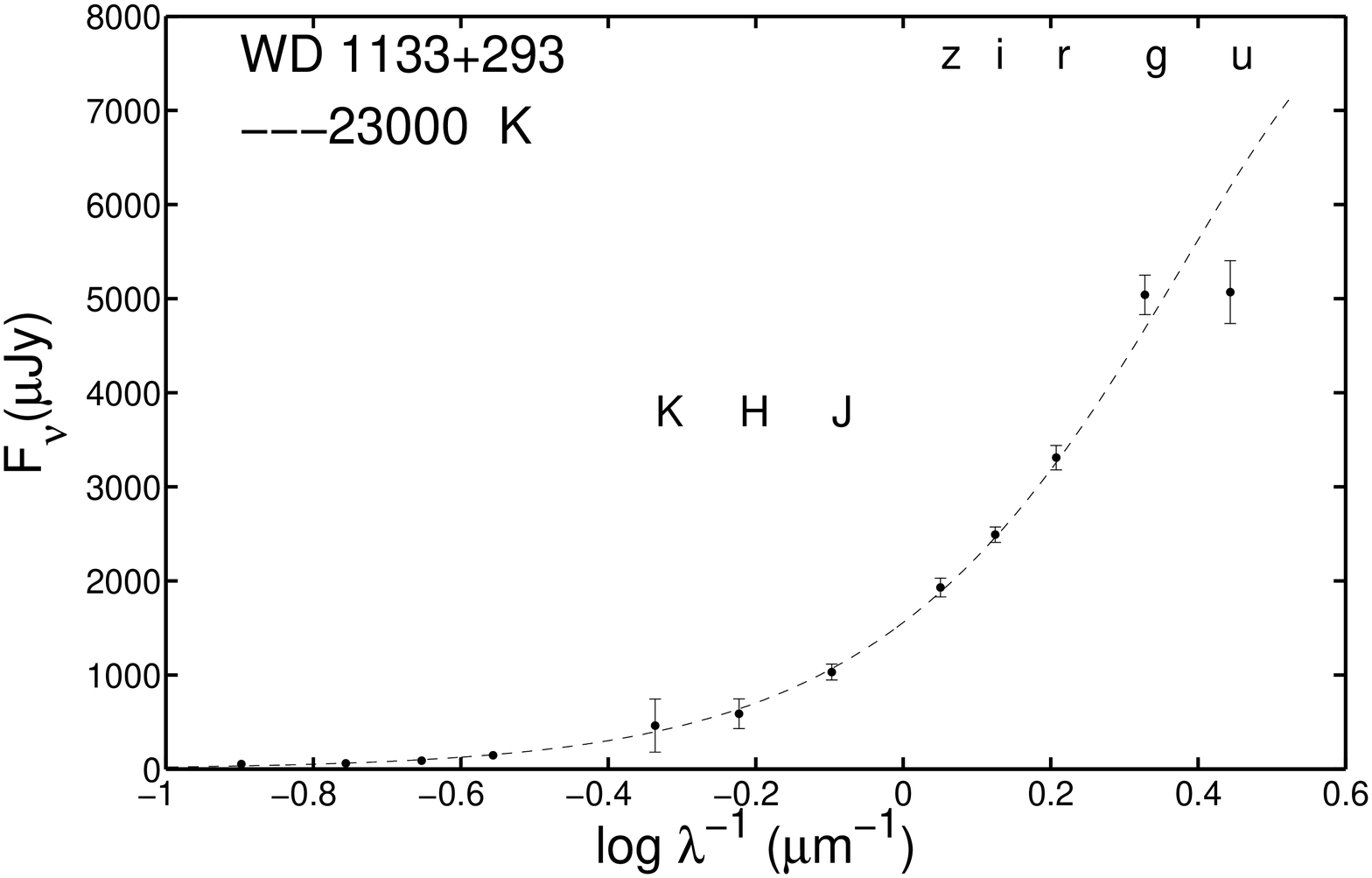}{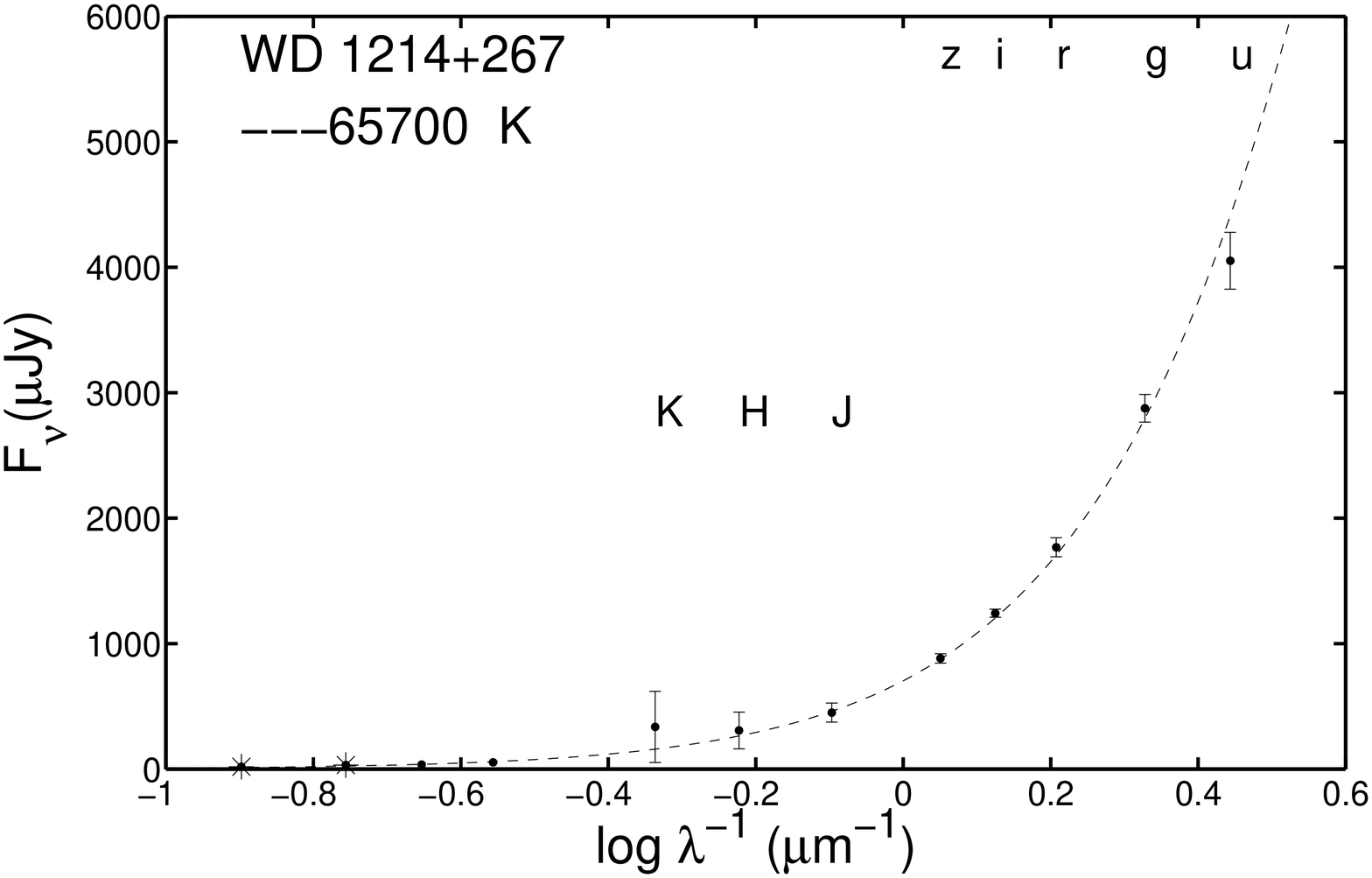}
\plottwo{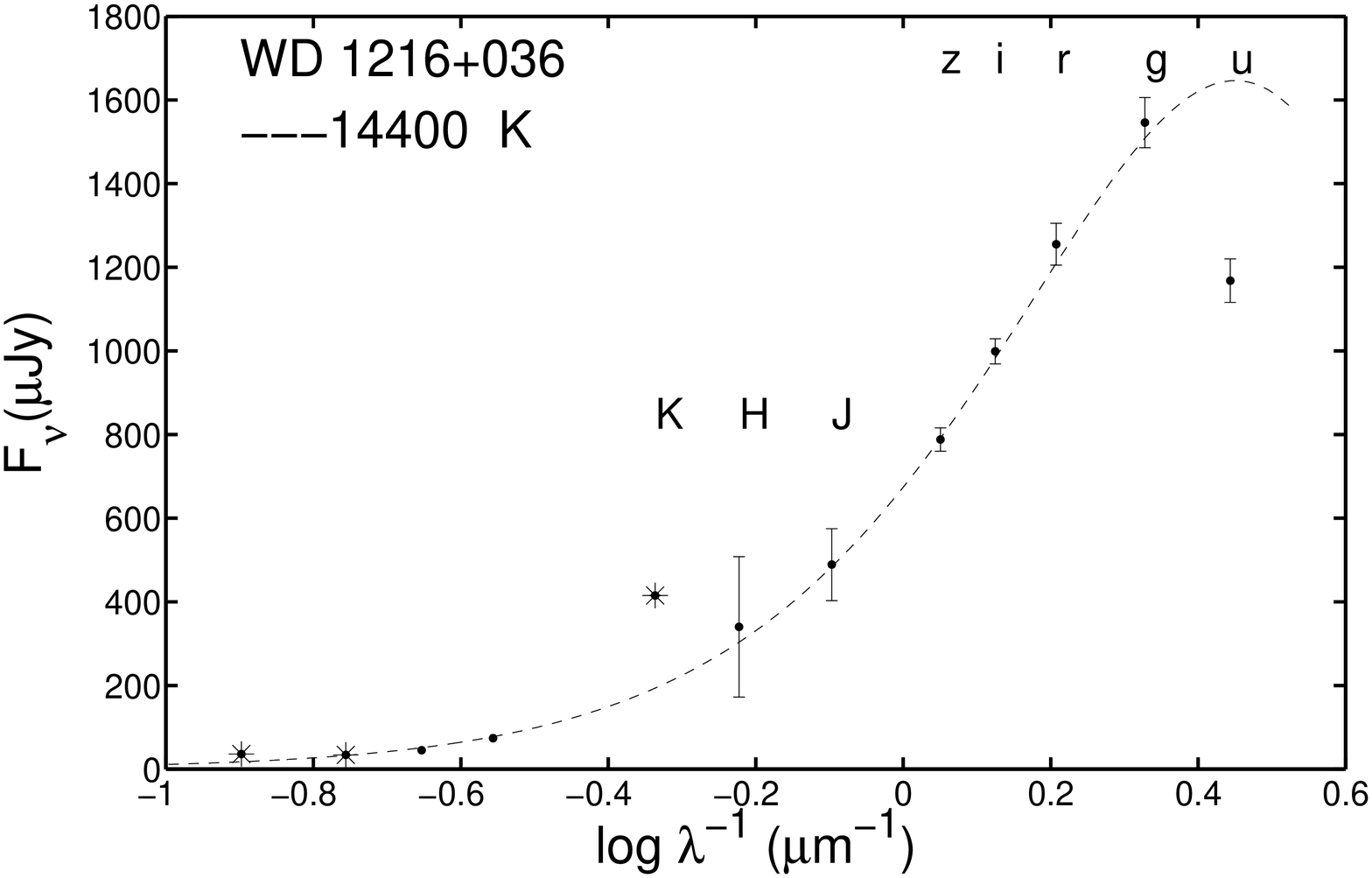}{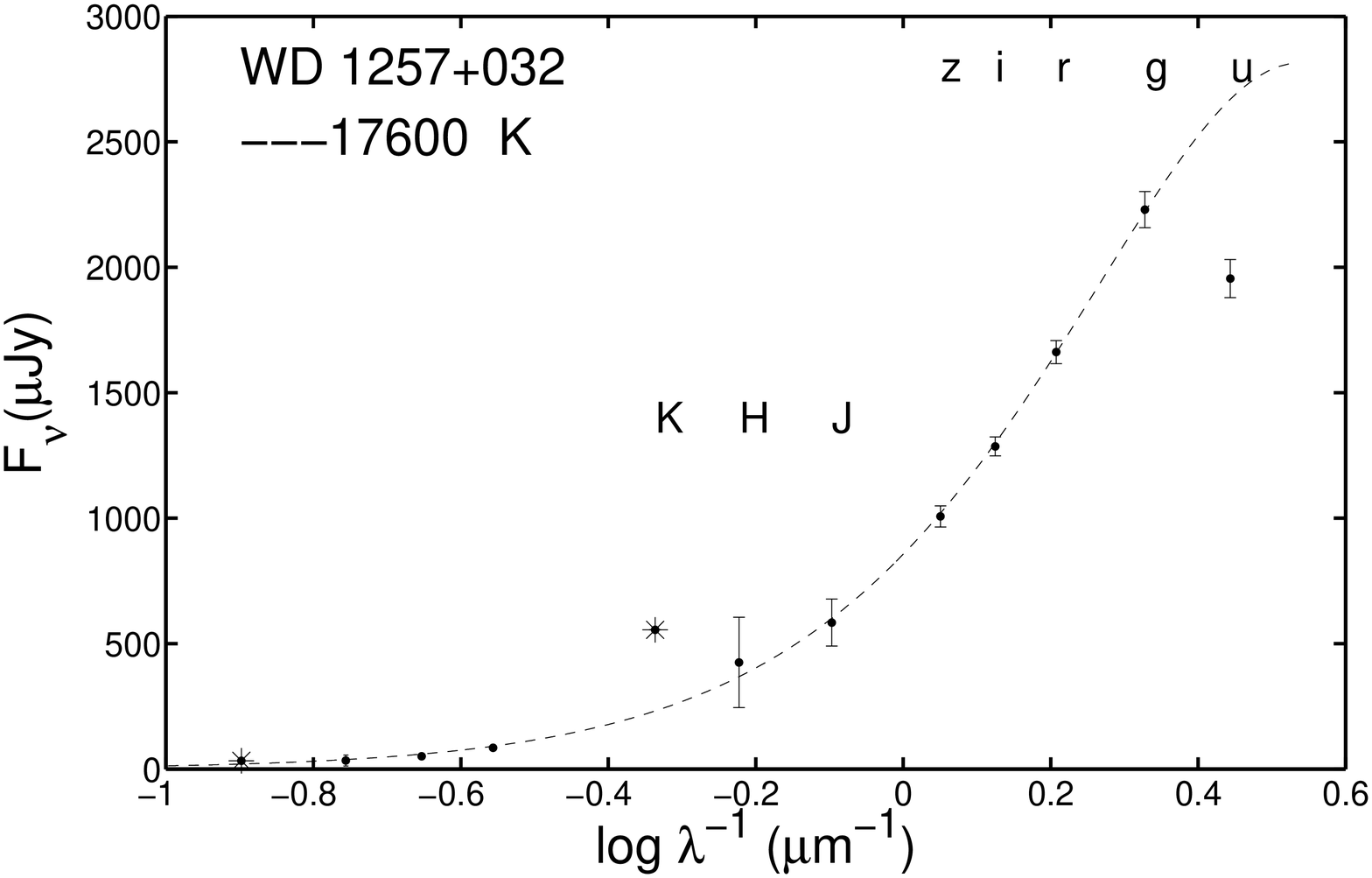}
\plottwo{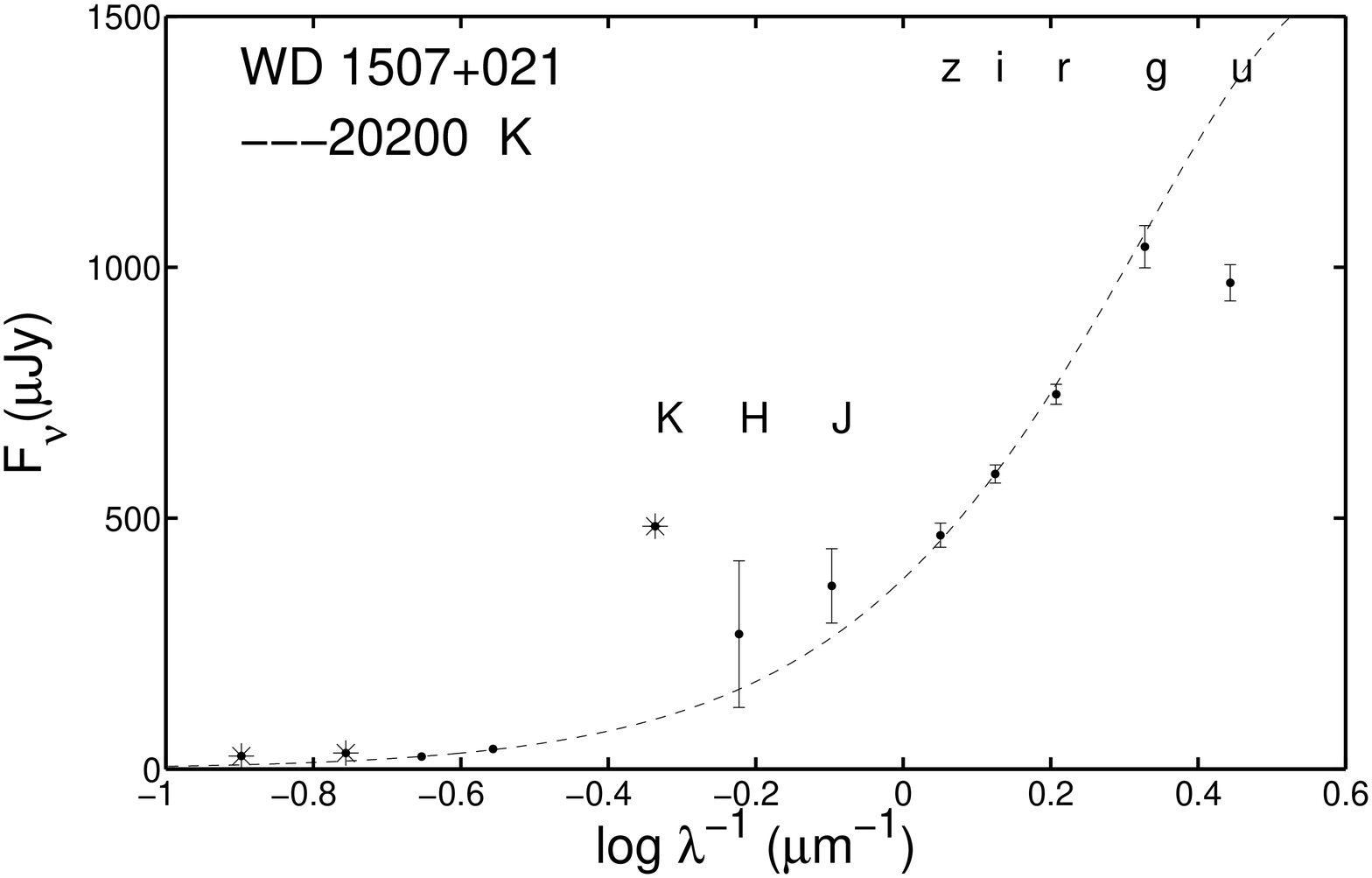}{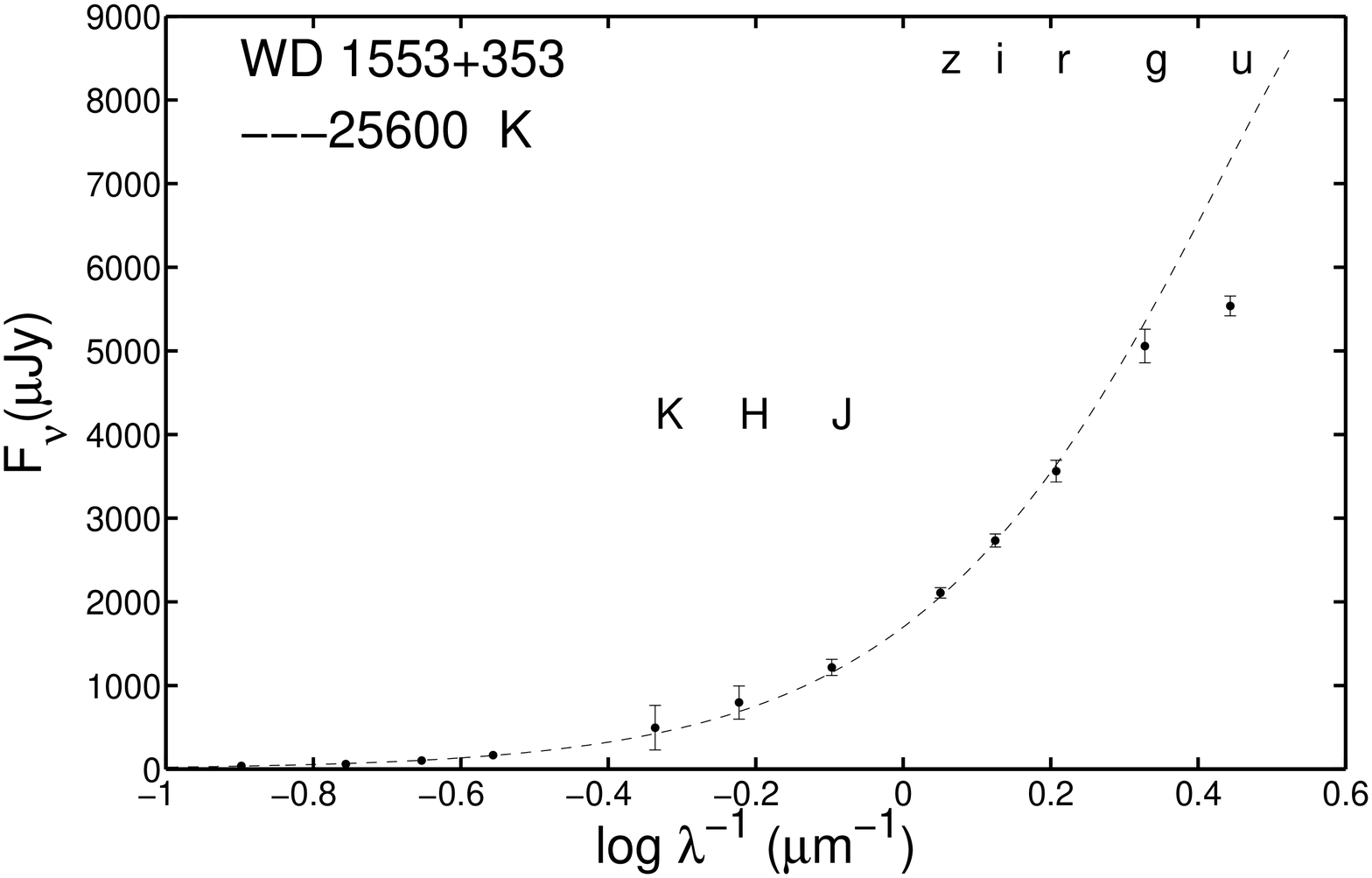}
\caption{The same as Figure \ref{Fig: ArchiveSED1}.} \label{Fig: ArchiveSED2}
\end{figure}

\begin{figure}
\epsscale{1.2}
\plottwo{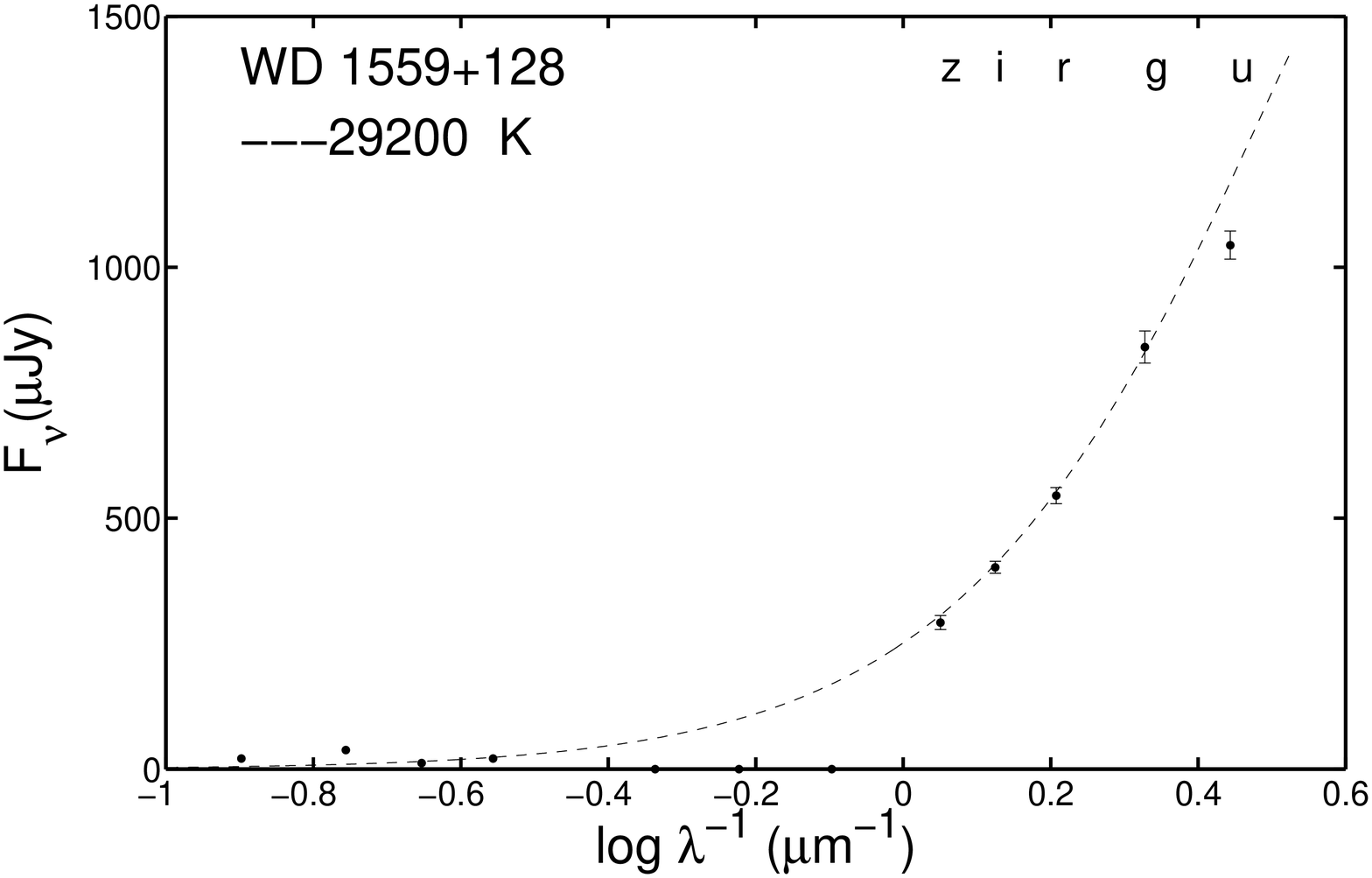}{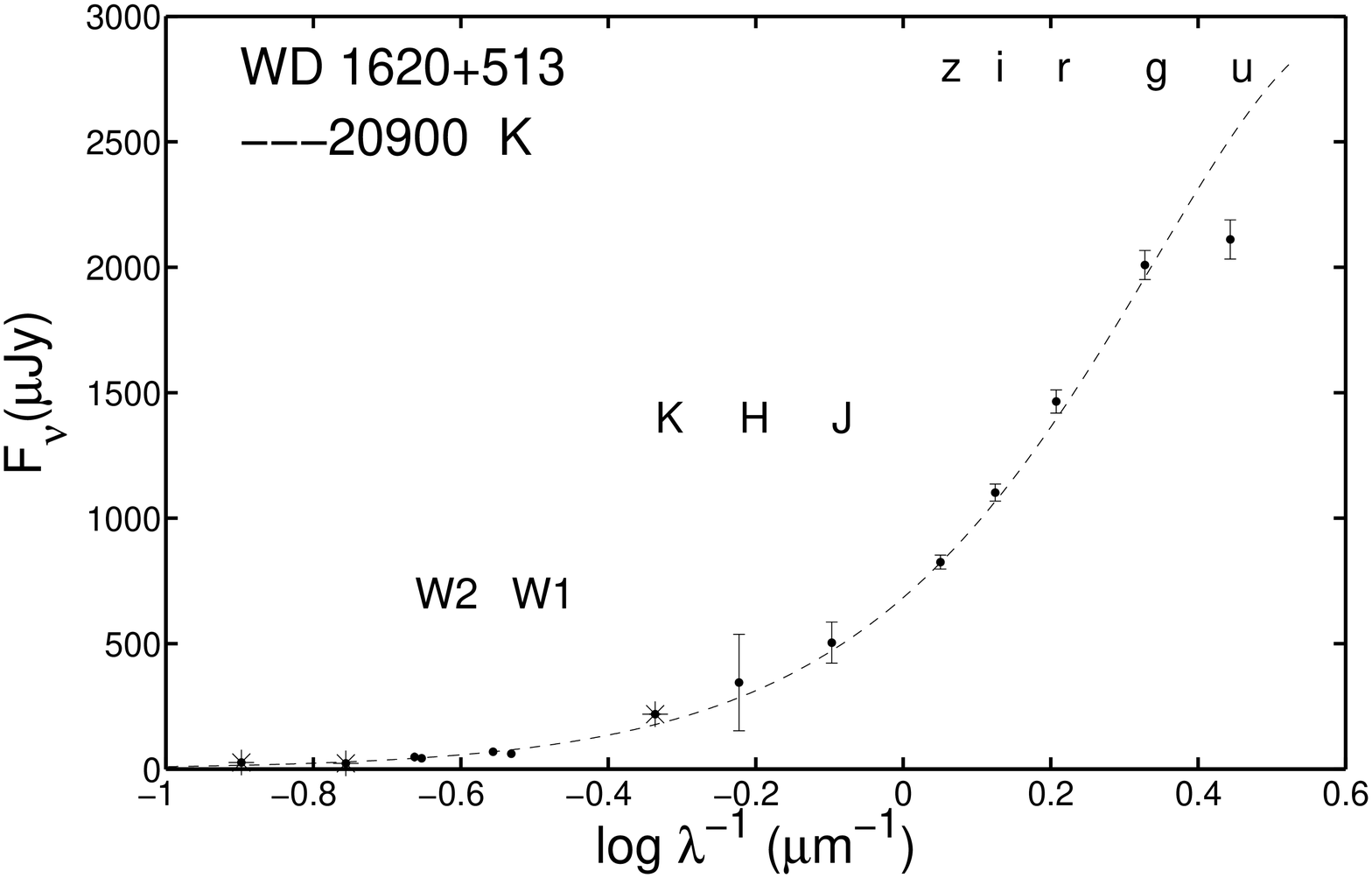}
\epsscale{0.55}
\plotone{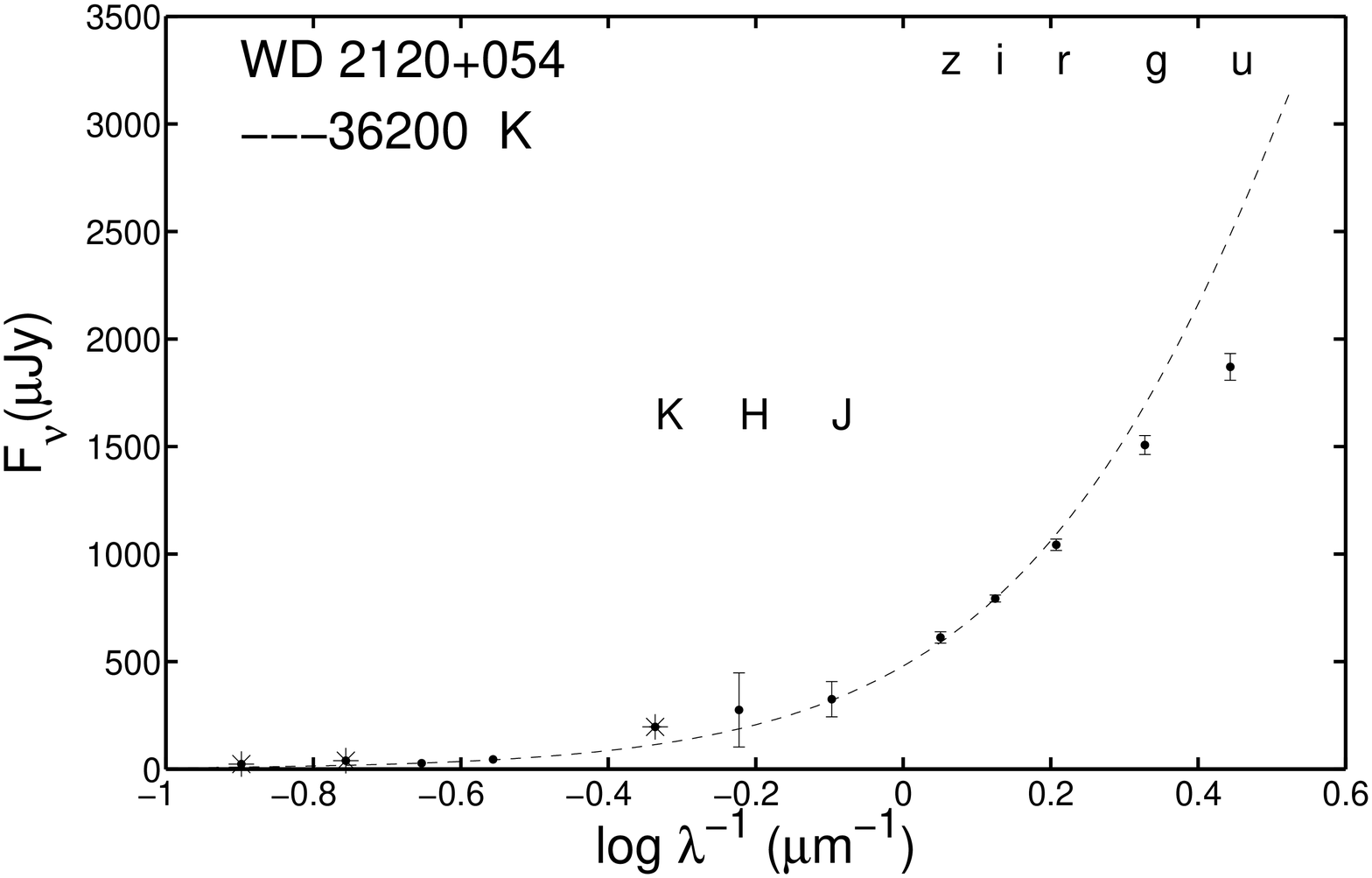}
\caption{The same as Figure \ref{Fig: ArchiveSED1}.} \label{Fig: ArchiveSED3}
\end{figure}

\begin{figure}[tbph]
\plotone{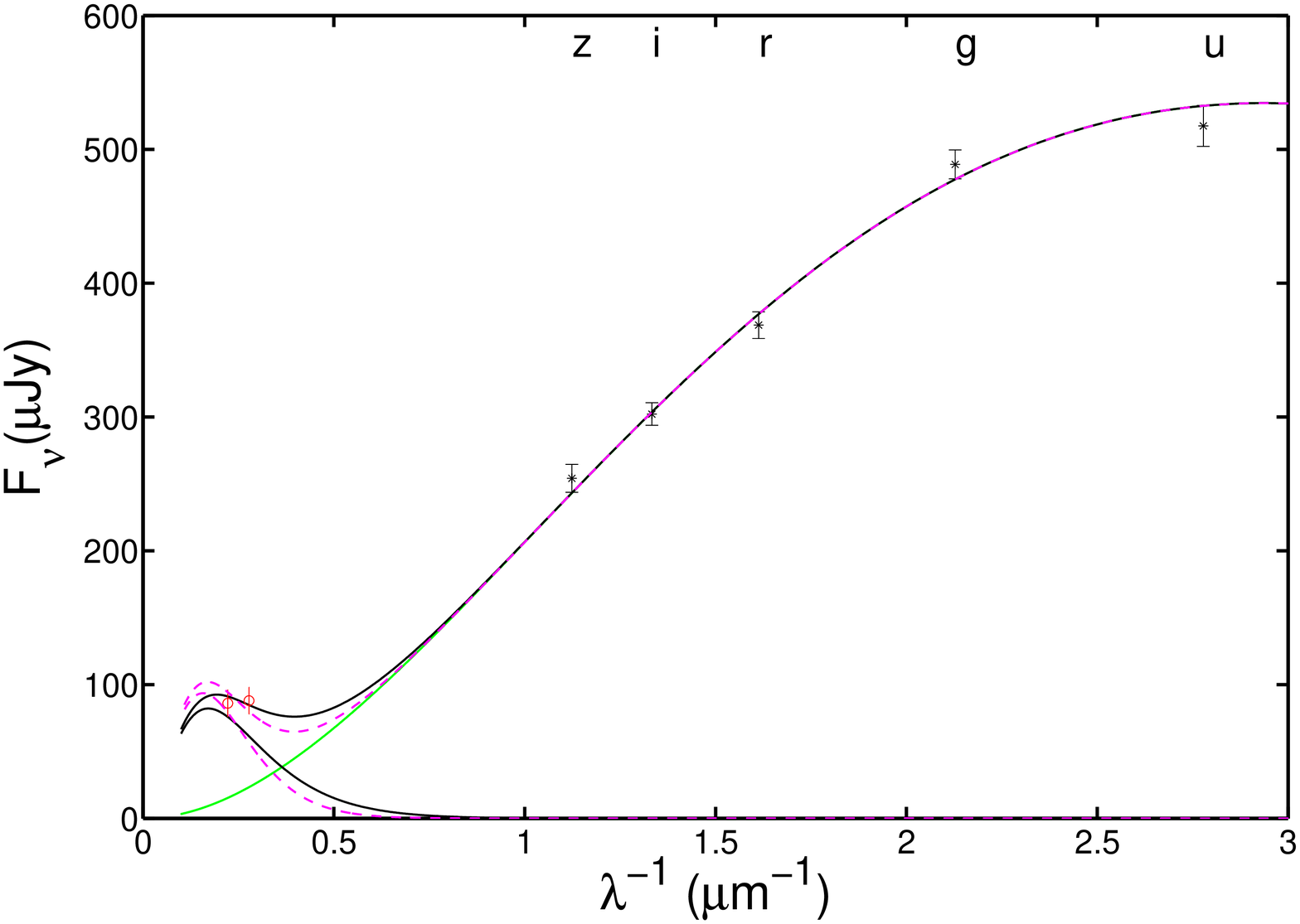}
\plotone{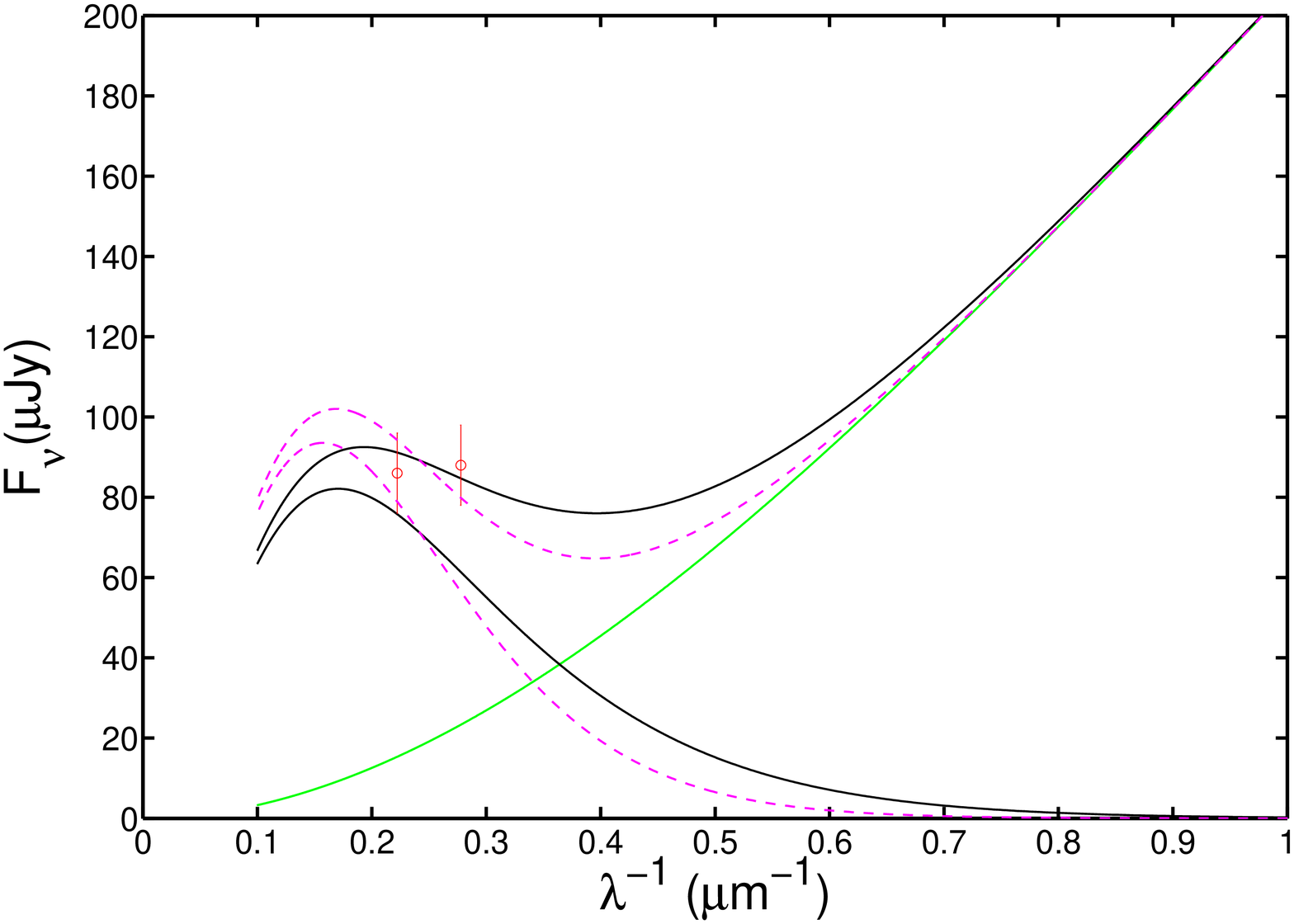}
\caption{SED for J2209+1223. The green line displays the photospheric flux from the star. The black lines and dashed pink lines represent the two disk models listed in Table \ref{Tab: DiskPara}: the flux from the disk and its total flux. The upper panel shows the fit to the entire model while the lower panel shows the fit to the disk; 2$\sigma$ error bars are displayed. \label{Fig: J2209}}
\end{figure}

\begin{figure}[tbph]
\plotone{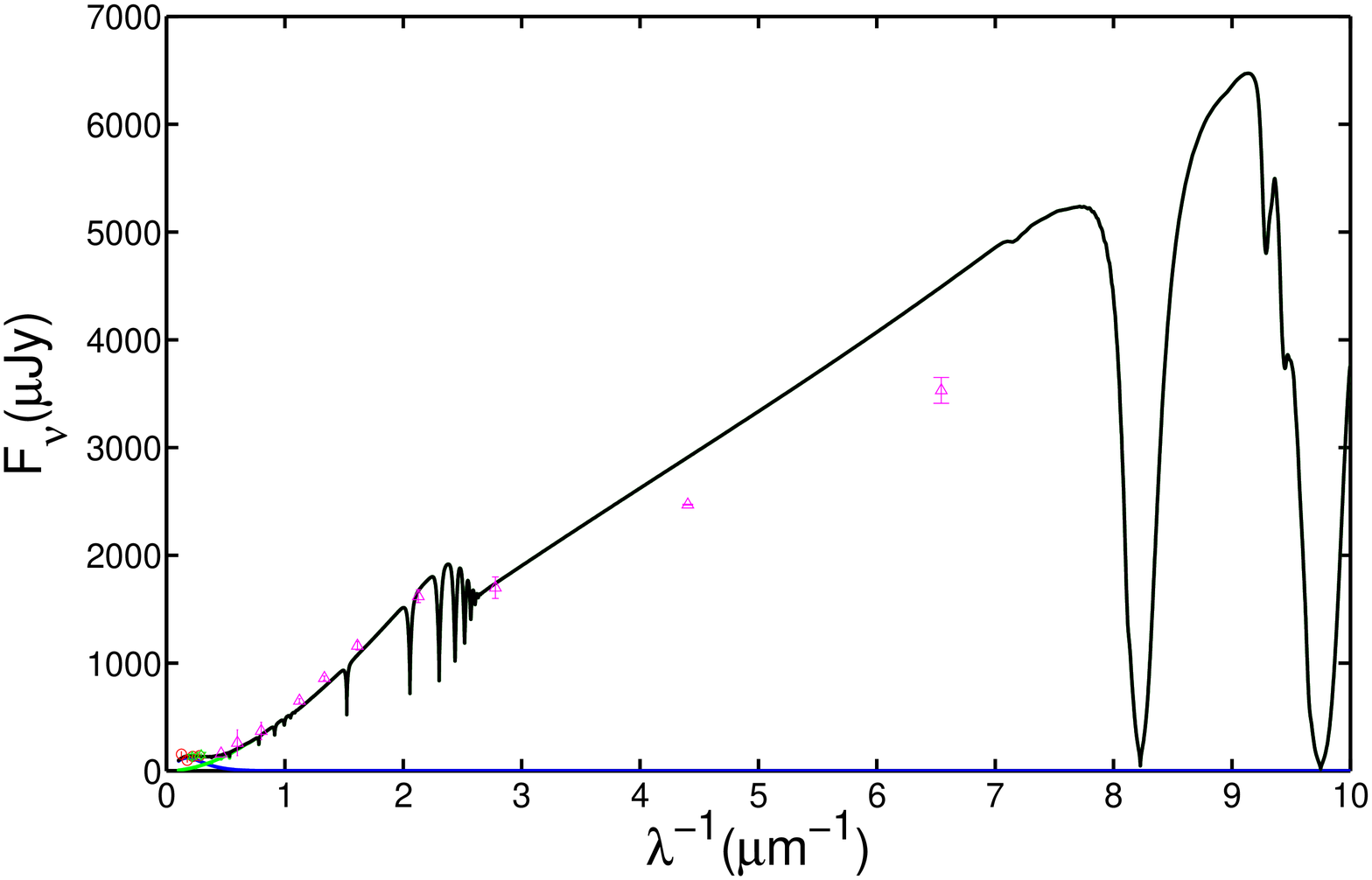}
\plotone{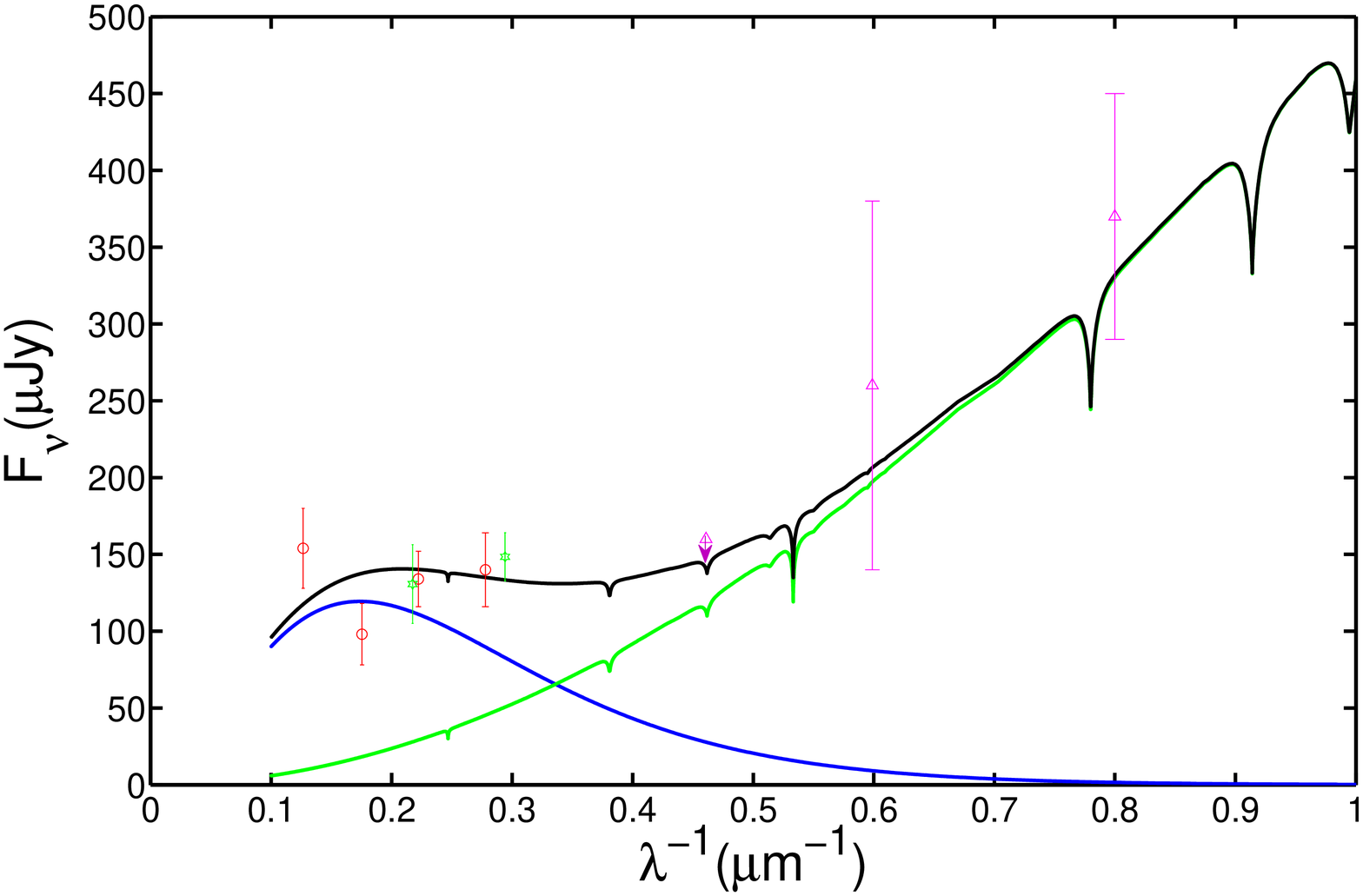}
\caption{SED for WD 0843+516. The green line displays the model atmosphere (D. Koester, private communication), the blue line the model disk and the black line the sum of both. The upper panel shows the entire fit to the model atmosphere while the lower figure only shows the disk portion. The pink dots are taken from GALEX, SDSS, 2MASS, green dots from WISE and red dots from IRAC, respectively; 2$\sigma$ error bars are displayed. \label{Fig: PG0843}}
\end{figure}

The Spectral Energy Distribution (SED) of all the targets are plotted in Figures \ref{Fig: SED7a} -- \ref{Fig: PG0843}. Also included in the SEDs are ugriz fluxes from the SDSS, JHK fluxes from the Two Micron All Sky Survey (2MASS) and WISE fluxes when the data are available. Ultraviolet fluxes from the {\it Galaxy Evolution Explorer} (GALEX) \citep{Martin2005} are usually excluded because they can be strongly suppressed due to heavy element blanketing \citep{Koester2011} or interstellar extinction. A blackbody model is then adopted to fit the star's photospheric flux, with most weight given to the SDSS photometry because the 2MASS fluxes have larger uncertainties. Though the blackbody temperature does not always agree with the reported white dwarf temperature of \citet{Dufour2007}, this method is sufficient to identify infrared excess that is 10\% above the photospheric value. When there is a detected excess, a thin, opaque, passive dusty disk is used to fit the SED \citep{Jura2003}. 

\subsection{Notes on Individual Stars}

\subsubsection{WD 0216-095}
This star is heavily blended with a nearby galaxy SDSS J021836.98-091955.7 with an angular separation of 2{\farcs}6 at position angle 210$^\circ$. We take a proper motion of 130 mas/yr in RA and 43 mas/yr in Dec from the NOMAD catalog \citep{Zacharias2005} to derive a position of 02:18:36.77-09:19:44.24 at the epoch of the {\it Spitzer} observation. Then we used the method described above to resolve these two objects. No excess is detected for this star.

\subsubsection{WD 0936+560}

In Figure \ref{Fig: SED7a}, we see the 2MASS H band upper limit is significantly lower than our predicted photospheric flux. Based on the SDSS photometry and the near infrared spectra for this star in \citet{Kilic2008}, we assign a very low weight for this point when fitting the SED and no infrared excess is found for this star.

\subsubsection{J2209+1223}
\begin{figure}[tbph]
\plotone{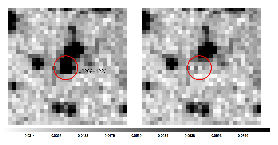}
\plotone{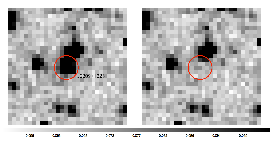}
\caption{IRAC mosaics of J2209+1223 in 3.6 $\mu$m (the upper panel) and 4.5 $\mu$m band (the lower panel) in the plate scale of 1{\farcs}2 pixel$^{-1}$; the left columns are the original data and the right columns are the residue after PRF fitting and the white dwarf is subtracted. North is up and East is left; the field of view is 36{\farcs}0 by 36{\farcs}0. The red circle is centered on J2209+1223 with a radius of 3 pixels and the nearby star is 6{\farcs}2 away at position angle 333$^\circ$. \label{Fig: PSF}}
\end{figure}

\begin{figure}[tbph]
\plotone{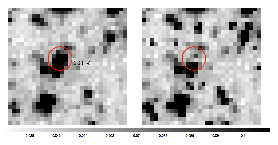}
\plotone{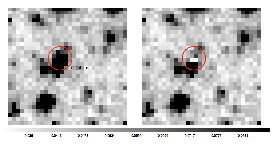}
\caption{The same as Figure \ref{Fig: PSF} except for G 241-6. We see this is a very complicated field and G 241-6 is heavily blended with an object with an angular separation of only 2{\farcs}3 at position angle 170$^\circ$ and another galaxy, J222334.23+683723.8 with a separation of 5{\farcs}5 and position angle 110$^\circ$. \label{Fig: psf_G241-6}}
\end{figure}

The IRAC image of this star reveals an unresolved background star separated at 6{\farcs}2 at position angle 333$^\circ$ (see Figure \ref{Fig: PSF}). The proper motion of this star is small enough that it is neglected. Robust centroid methods were used in each exposure to accurately determine the position of these two objects, and then PRF fitting was performed. To demonstrate the effectiveness of the PRF fitting, the original image and residue after subtracting the star are shown in Figure \ref{Fig: PSF}.

Figure \ref{Fig: J2209} shows the SED for J2209+1223; we see a strong infrared excess in both IRAC bands, which is better explained by a flat disk rather than a companion. In the disk models of \citet{Jura2003}, there is a degeneracy between the inclination of the disk and its size. With only two data points, there are two sets of parameters as listed in Table \ref{Tab: DiskPara} that can reproduce the SED: a more inclined larger disk or more face-on smaller disk. Another good fit to the SED is a blackbody with T = 1000 K and R = 9R$_{\odot}$ at the same distance as the white dwarf, 190 pc. However, this derived radius is much too large for a brown dwarf. Furthermore, there is no feature indicative of M dwarf companions, such as H$\alpha$ emission or TiO absorption lines in the SDSS spectra of this star.

J2209+1223 has an effective temperature of 17,300 K with an accretion rate of 3 $\times$ 10$^{10}$ g s$^{-1}$; this value is quite uncertain as it is derived from scaling the equivalent width of magnesium lines from GD 61, which is a white dwarf with similar stellar parameters \citep{Farihi2011a}. This star falls into the catalog of highly polluted warm white dwarfs with an infrared excess.

\subsubsection{G 241-6}

We take the proper motion of G 241-6 as 144 mas/yr in RA and 244 mas/yr in Dec \citep{Zacharias2005} to derive its position at 22:23:33.38+68:37:26.77 in the {\it Spitzer} observation. In Figure \ref{Fig: psf_G241-6}, we can see in both IRAC bands this star is heavily blended with an unknown object at an angular separation of 2{\farcs}3 at position angle 170$^\circ$. Another galaxy is present at J222334.23+683723.8 at an angular distance of 5{\farcs}5 and position angle 110$^\circ$. By using PRF fitting, we successfully resolve these objects and no excess is found even though this 15,300 K star is as heavily polluted as GD 40 \citep{Zuckerman2010}, which does have a dust disk \citep{Jura2007b}. 

\subsubsection{WD 0819+363}

In four IRAC bands, this star is blended with the galaxy SDSS J082245.87+361410.2 with a distance of 3{\farcs}5 at position angle 240$^\circ$ and another unknown object separated at 4$\farcs$1 and position angle 110$^{\circ}$. The proper motion of this star is -78 mas/yr in RA and -58 mas/yr in Dec \citep{Zacharias2005}, and it is located at 8:22:46.2+36:14:12.04 at the epoch of the {\it Spitzer} observation. PRF fitting was performed in the 3.6 $\mu$m and 4.5 $\mu$m band and no excess was found. In the 5.7 $\mu$m and 7.9 $\mu$m band, the background is so noisy that APEX failed to perform PRF fitting in the nominal position derived from IRAC band 1 and 2. Instead we used aperture photometry at the position of the star and there appears to be a 3$\sigma$ excess as shown in Figure \ref{Fig: ArchiveSED1}. However, this is unlikely to be real considering the complicated field.

\subsubsection{WD 0843+516}

Fig \ref{Fig: PG0843} shows the SED for WD 0843+516, apparently the hottest white dwarf so far detected to have a dust disk. The model atmosphere was provided by D. Koester (private communication). An opaque disk model can reproduce the data with the parameters listed in Table \ref{Tab: DiskPara}. As previously found with GD 362 \citep{Jura2007b}, the model tends to underpredict the flux at 7.9 $\mu$m, which might be contaminated by strong silicate emission in this band. An M dwarf companion can be ruled out as there is no excess in the JHK bands (see \citet{Hoard2007}). Another fit to the SED can be achieved with a blackbody of T = 800 K and R = 0.22R$_{\odot}$ at 80 pc. However, this radius is much too large for a brown dwarf, so this possibility is also excluded. So far WD 0843+516 is the only white dwarf with an infrared excess that is yet to know whether it is polluted as there is no suitable spectra reported in the literature.

\section{RESULTS}

\subsection{Missing Planetary Debris Around Cool Stars}

Warm dust has been found around 21 white dwarfs with temperatures ranging from 9,500 K to 24,000 K and at least 20 of them have a highly polluted atmosphere. Between 1\% to 3\% single white dwarfs with cooling ages less than 0.5 Gyr possess warm circumstellar dust \citep{Farihi2009}. In contrast, the disk fraction around cool white dwarfs (T$_*$ $<$ 6000 K) only has an upper limit of 0.8\% \citep{Kilic2009a}. Previously, 11 DZs have been targeted with {\it Spitzer}/IRAC \citep{Farihi2008a, Farihi2008b, Farihi2009}, but none shows an infrared excess. In this study, we observed 10 additional DZs, which are heavily polluted and most likely disk-host stars. Though almost doubling the numbers of DZs, we still fail to find any infrared excess. We address the implications and possible explanations in section 5 and 6.

\subsection{Hot White Dwarfs with a Close-in Disk \label{sec: HotWDDisk}} 

As discussed by \citet{VonHippel2007}, we need to understand how solid dust grains can survive around hot white dwarfs. Dust disks have been confirmed around four stars with stellar temperature higher than 20,000 K: WD 0843+516 (this paper), GALEX 1931 \citep{Debes2011}, PG 1457-086 \citep{Farihi2009} and WD 1226+110 \citep{Brinkworth2009}. Currently, there are two different models to describe the inner boundary of the disk: a fully opaque disk model \citep{Jura2003} and an opaque disk with a thin transition zone, which goes from optically thin to optically thick \citep{Rafikov2011a}. However, for the hot white dwarfs, only the opaque disk model allows the inner boundary of the disk to be within the tidal radius,  

\begin{equation} \label{equ: rs}
r_{inner}=\left( \frac{2}{3\pi} \right)^{1/3}\left( \frac{T_*}{T_s} \right)^{4/3}r_*
\end{equation}
where T$_*$  and r$_*$ is the stellar temperature and radius. T$_s$ is the temperature at which the dust particles sublimate; it can be constrained from the 3.6 $\mu$m excess and it is usually 1200 K (see, for example, \citet{Jura2007b}). So for a 20,000 K star, the inner boundary for a complete opaque disk is 25 r$_*$.

In the model described by \citet{Rafikov2011a}, the inner boundary of the optically thin region is,

\begin{equation} \label{equ: rss}
r_{inner}=\frac{r_*}{2}\left( \frac{T_*}{T_s} \right)^2
\end{equation} 
For a 20,000 K star, the inner radius is 140 r$_*$, whereas the tidal radius is approximately 130 r$_*$ \citep{VonHippel2007}. In this case, all the dust particles lie outside of the tidal radius.

We have no way to resolve the disk and directly tell which model is correct, but the fully opaque disk model is the simplest explanation for the presence of close-in dust. There are two additional arguments in favor of the opaque disk model: it can reproduce the substantial amount of infrared excess and the dust disk radii agree with those derived from gaseous emission lines \citep{Melis2010}. Further discussion is in presented in the Appendix, where we consider P-R drag on a completely opaque disk described by \citet{Jura2003} and derive an accretion rate that agrees better with the data than the model described in Equation (\ref{equ: rss}) \citep{Rafikov2011a}.

\section{CHARACTERISTICS OF DISK-HOST STARS}

We see the planetary debris seems to be missing around the polluted cool stars. Why is that? In this section, we explore different characteristics that might be correlated with the occurrence of a dust disk.

\subsection{High Heavy Element Accretion Rate?}

\begin{figure}[tbph]
\plotone{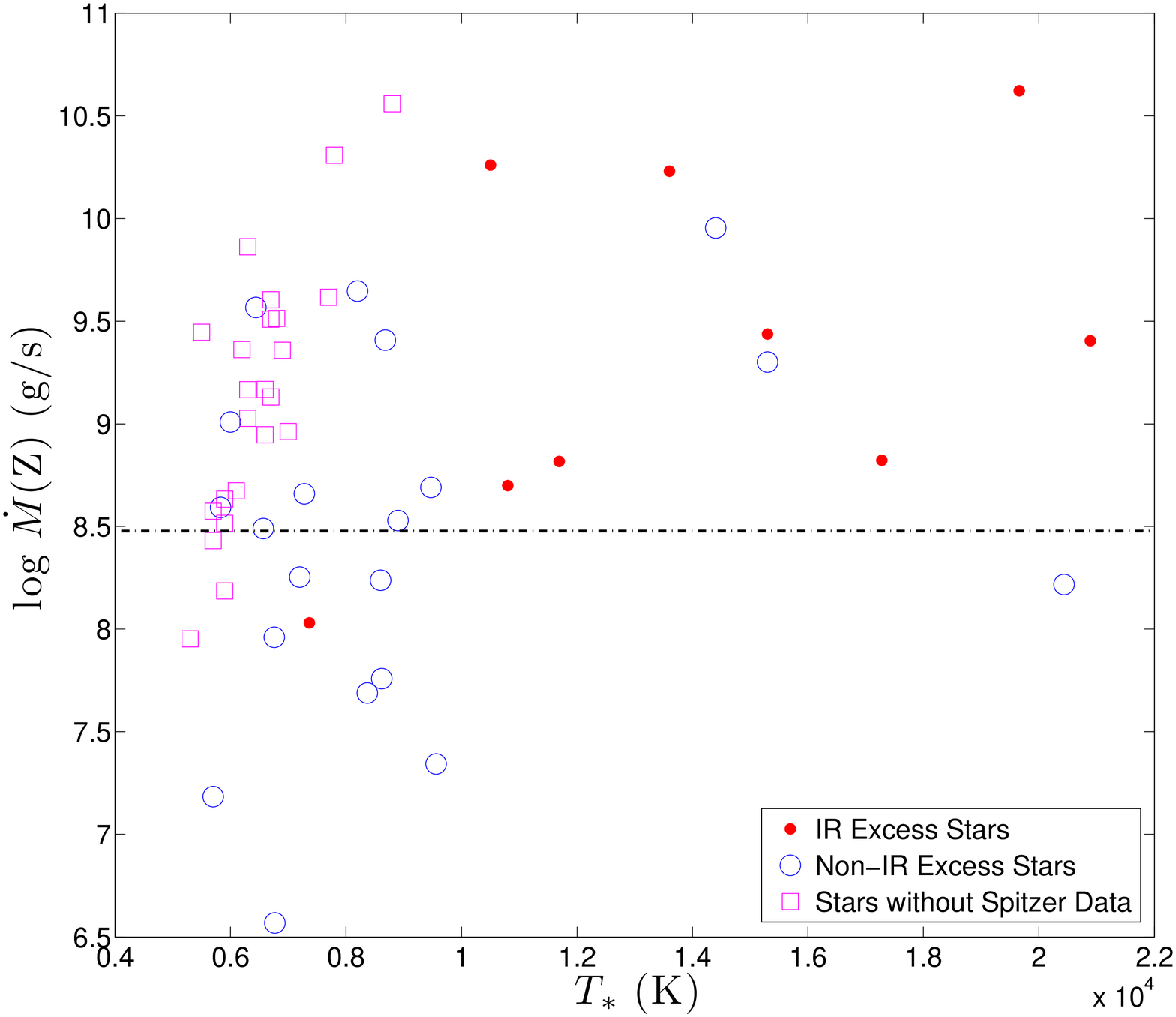}
\caption{Accretion rates at different stellar temperatures. The data include stars with well determined major element abundance, newly identified heavily polluted DZs \citep{Koester2011}, DAZs \citep{Zuckerman2003} and Cycle 7 stars reported in this paper. The dash-dotted line represents an accretion rate of 3 $\times$ 10$^8$ g s$^{-1}$, a threshold proposed by \citet{Farihi2009}: over 50\% of single white dwarfs with accretion rate at least this high display an infrared excess from $\sim$1000 K dust.}
\label{Fig: dMdt}
\end{figure}

It is found that at least 50\% of white dwarfs with heavy element accretion rate over 3 $\times$ 10$^8$ g s$^{-1}$ have warm dust \citep{Farihi2009}. But cool stars do not appear to follow this pattern. Figure \ref{Fig: dMdt} shows a comparison of the overall accretion rate with a white dwarf's stellar temperature. We see that there are numerous cool white dwarfs (T$_*$ $<$ 10,000 K),  that have an accretion rate higher than 3 $\times$ 10$^8$ g s$^{-1}$, but none displays a dust disk. 

To calculate the accretion rate of element A, we assume a steady state so it equals the mass of A in the convective zone divided by its settling time. For the stars that have well determined major element abundances, the accretion rate is simply the sum of all the heavy elements. For the stars that have both magnesium and iron abundance, we assume they are 30\% of the total mass of the parent body. For the Cycle 7 targets, calcium is the only detected element and we assume it is 0.6\% of the total mass, extrapolated from the cool DZs in \citet{Koester2011}. All the stellar parameters are extrapolated from \citet{Koester2009a, Koester2011}.

\subsection{Massive Parent Bodies?}

\begin{figure}[tbph]
\plotone{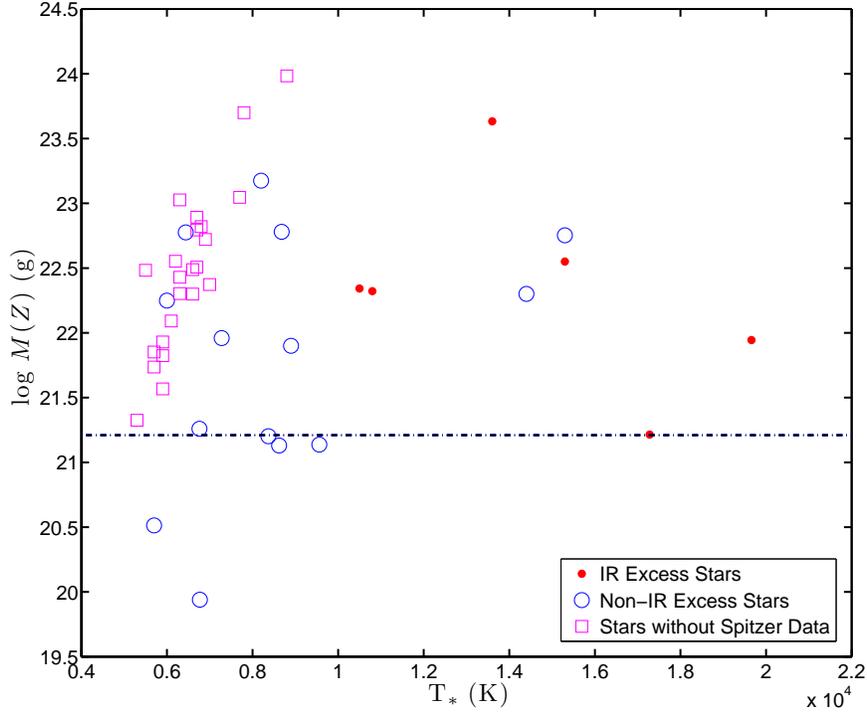}
\caption{Mass of the pollution in the convective zone versus temperature for all the DBZs and DZs in Figure \ref{Fig: dMdt}. The black dashed line denotes the mass in the mixing zone of GD 61, 1.6 $\times$ 10$^{21}$g, which is the minimum mass found so far to have an infrared excess \citep{Farihi2011a}. We see most of the stars are accreting from parent bodies more massive than this value. \label{Fig: MT}}
\end{figure}

Another possibility is that while more massive parent bodies can form a dusty disk, less massive parent bodies can only form a more tenuous, short-lived dust disk or a purely gaseous disk (see \citet{Jura2008}). A comparison of the mass of heavy elements in the atmosphere versus the white dwarf's effective temperature is shown in Figure \ref{Fig: MT}. We see there is no difference in the mass in the convective zone, $M(Z)$, between stars with and without an infrared excess. An uncertainty is that the mass plotted in Figure \ref{Fig: dMdt} are only lower bounds since we do not know how long the accretion has been going on. Also, there is no reason to think that the as-yet-to be accreted mass in the disk is a function of the star's temperature.

\subsection{Refractory-rich Parent Bodies?}

\begin{figure}[tbph]
\plotone{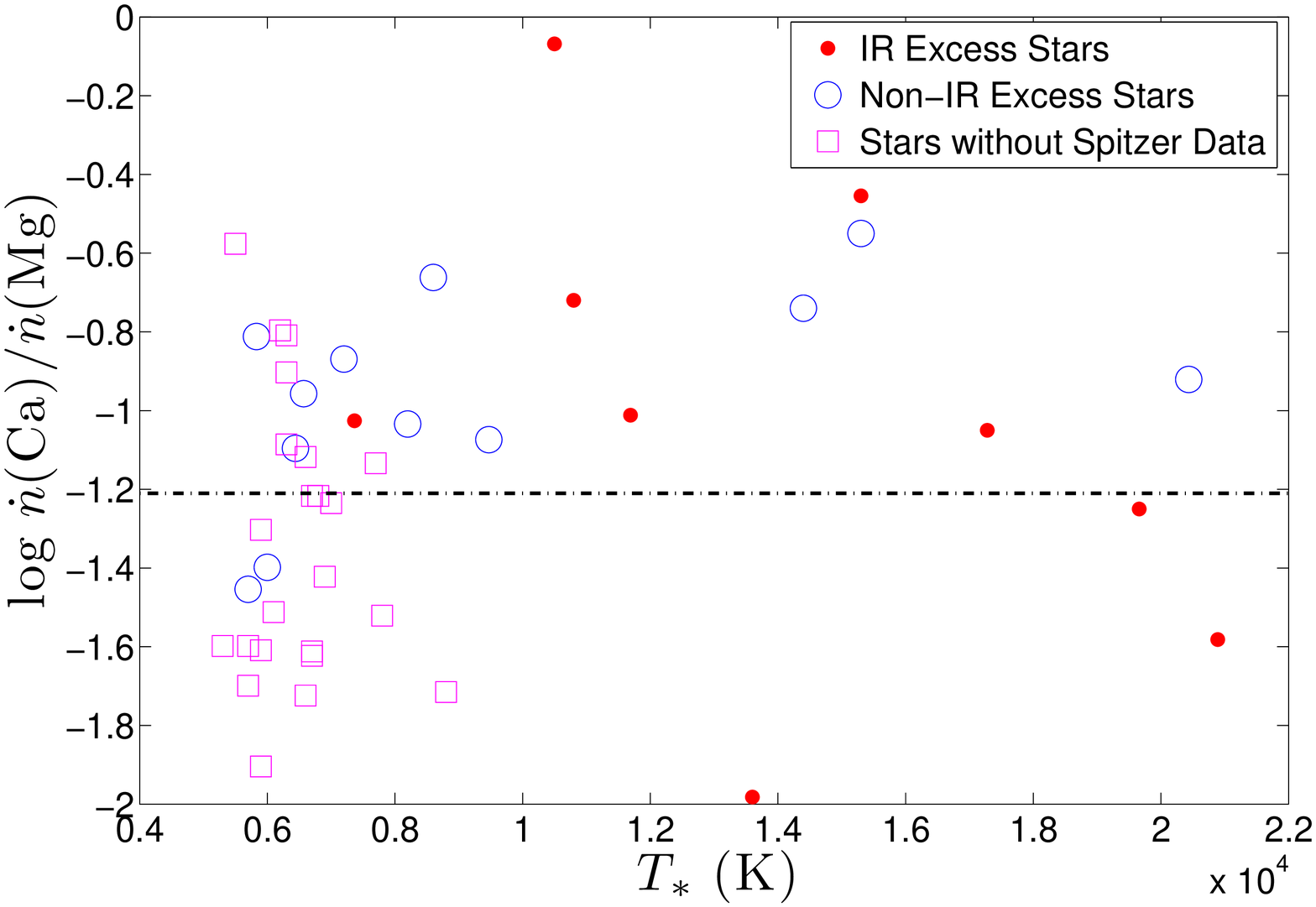}
\plotone{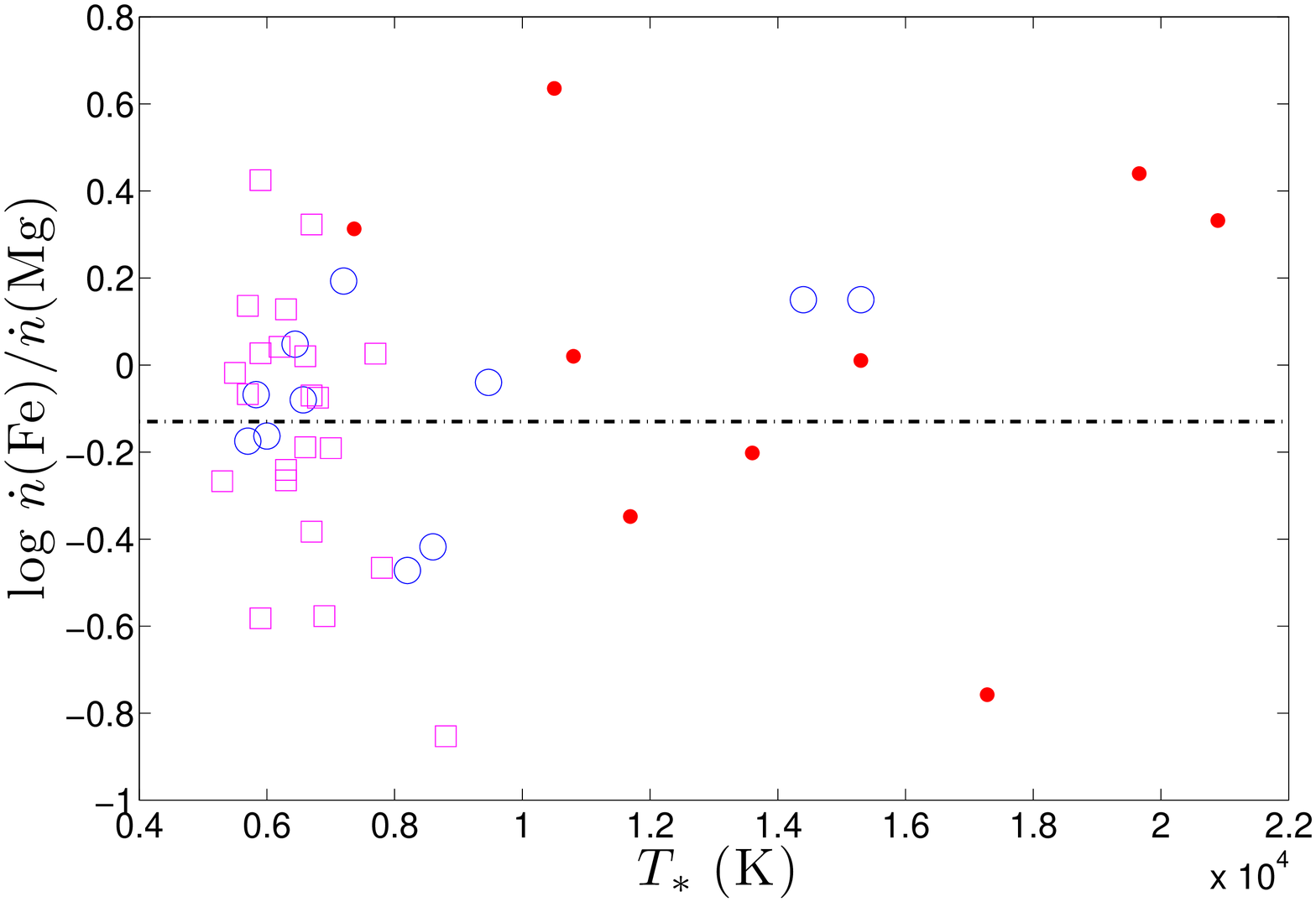}
\caption{Compositional difference for all the stars in Figure \ref{Fig: dMdt} that have Ca, Mg and Fe detections. The relative abundance has been corrected for settling assuming a steady state. The dashed line is the value for bulk Earth. We see the stars with infrared excess usually have a high $\dot{n}$(Ca)/$\dot{n}$(Mg). \label{Fig: CaMgFe}}
\end{figure}

\citet{JuraXu2012} argue that there are at least two populations of parent bodies that are accreted onto a white dwarf: water-rich comet-like objects and dry rocky objects. Here we suggest that the compositional variations of the accreted materials might be important in the formation of a dust disk. A comparison between calcium, magnesium and iron accretion rates is plotted in Figure \ref{Fig: CaMgFe}. log $\dot{n}$(Ca)/$\dot{n}$(Mg) and log $\dot{n}$(Fe)/$\dot{n}$(Mg) versus the star's temperature are shown in order to correct for different settling rates. These values correspond to the intrinsic abundance ratio in the parent body assuming a steady state. 

We see in the upper panel of Figure \ref{Fig: CaMgFe} that most stars with an infrared excess have relatively high abundance of Ca, a highly refractory element \citep{Allegre2001}; the average [Ca]/[Mg] in these stars is -0.71 while it is -1.21 in bulk Earth. In contrast, when measured in a different sample of DZs than studied here, [Ca]/[Mg] equals -1.42 \citep{Koester2011}. In the lower panel of Figure \ref{Fig: CaMgFe}, we see Fe abundance almost stays the same. Since iron is one of the dominant elements in bulk Earth as well as white dwarf pollution \citep{Klein2010}, there is no reason to think its abundance should vary a lot. There is additional evidence supporting that cool white dwarfs are accreting more volatile-rich materials. Sodium, which is an important volatile, has a nearly solar abundance in 28 extreme DZs, much higher than the bulk Earth value \citep{Koester2011, Allegre2001}. 

\section{ORIGIN OF HEAVY ELEMENTS IN STARS WITHOUT AN INFRARED EXCESS}

We see most polluted white dwarfs do not possess a dust disk. So where do the heavy elements come from? Interstellar accretion was traditionally considered to be the origin of this material \citep{Dupuis1993a} but this model has faced many challenges \citep{Farihi2010a, Koester2011}. We discuss two avenues of accretion that can pollute a white dwarf's atmosphere without producing a dust disk.

{\it Accretion from a distant reservoir.} It has been suggested that some comets go through natural fragmentation due to fast rotation \citep{Drahus2011} and in this way contribute to the zodiacal light in our solar system \citep{Nesvorny2010}. A similar process might happen around a white dwarf given that there might be a large reservoir of comets. Eventually some of the dust particles will drift inward due to P-R drag and the amount of infrared flux produced at frequency $\nu$ is \citep{Jura2007b},

\begin{equation} \label{equ: Fnu}
F_{\nu} \approx \frac{1}{2}ln\left(\frac{R_{i}}{R_{f}}\right) \frac{\dot{M}(Z)c^2}{\nu}/4\pi D^2
\end{equation}
where R$_{i}$, R$_{f}$ is the initial and final distance of the accretion and R$_i$/R$_f$ is taken to be 1000, $\dot{M}$(Z) is the total accretion rate, c is the speed of light and D is the distance between the Sun and the white dwarf. We compare the flux predicted from Equation (\ref{equ: Fnu}) with our data for Cycle 7 targets in Table \ref{Tab: Comp}. For the heavily polluted Cycle 7 targets, we see that the calculated flux is always much higher than the observed value. However, there are other cool white dwarfs with an accretion rate $\sim$ 10$^6$ g s$^{-1}$ \citep{Farihi2009} where this process might be important.

{\it Accretion from an orbiting gaseous disk.} A gaseous disk that does not emit too much in the infrared can be formed if the grains mutually annihilate as described in the model of \citet{Jura2008} or the objects sublimate as they get close to the white dwarf. This is likely to be true for DAZs warmer than 11,000 K because their settling time is only days \citep{Koester2009a}. Without the continuous feeding of material, the elements would have quickly settled.

\section{CONCLUSIONS}

We find two new stars with an infrared excess, J2209+1223 and WD 0843+516, apparently the hottest white dwarf with a close-in disk. We fail to find any warm dust around heavily polluted DZs, all of which have T$_*$ $<$ 10,000 K. We raise the possibility that there might be some correlation between the occurrence of dust disk and the volatility of the accreted material. The best model to explain the source of pollution in white dwarfs without an infrared excess has yet to be established.

We thank J. Farihi for useful email exchanges on {\it Spitzer} data reduction and D. Koester for the computation of white dwarf model atmospheres. S. Xu thanks H. Meng and D. Rodriguez for helpful suggestions and discussions. This work has been supported by the National Science Foundation.

\appendix
\section{APPENDIX}

In this section, we extend the model of \citet{Rafikov2011a} for P-R drag on the entire disk to account for the observed mass accretion rates. Our modification of his model leads to a better agreement with the data.

Assuming a steady state, the accretion rate provided by P-R drag is \citep{Rafikov2011a},

\begin{equation} \label{Eq: 1}
\dot{M}(r_{inner})=\frac{16\phi_r}{3}\frac{r_*^3}{r_{inner}}\frac{\sigma T_*^4}{c^2}
\end{equation}
where $\sigma$ is the Stefan Boltzmann constant, c is the speed of light, and $\phi_r$ is an efficiency coefficient.

Using Equation (\ref{equ: rss}), \citet{Rafikov2011a} derived an accretion rate,

\begin{eqnarray} \label{MZr}
\dot{M}(r_{inner})	&=&	\frac{32\phi_r}{3}\sigma \left( \frac{r_*T_*T_s}{c} \right) ^2 \nonumber \\
			&=& 1.80 \times 10^8 \left( \frac{T_*}{15000K} \right)^2 g\; s^{-1} 
\end{eqnarray}
In the last step we take r$_*$=0.013 R$_{\odot}$ and $\phi_r$ = 1.

As discussed in section \ref{sec: HotWDDisk}, our inner boundary is described by Equation (\ref{equ: rs}); so we get a different value of accretion rate,

\begin{eqnarray} \label{MZ}
\dot{M}(r_{inner})	&=&  \left(\frac{3\pi}{2}\right)^{1/3}\frac{16\phi_r\sigma}{3} \left( \frac{r_*}{c}\right)^2 (T_sT_*^2)^{4/3}  \nonumber \\
			&=&  8.14 \times 10^8 \left(\frac{T_*}{15000K}\right)^{8/3} g\; s^{-1}
\end{eqnarray}

\begin{figure}[tbph]
\plotone{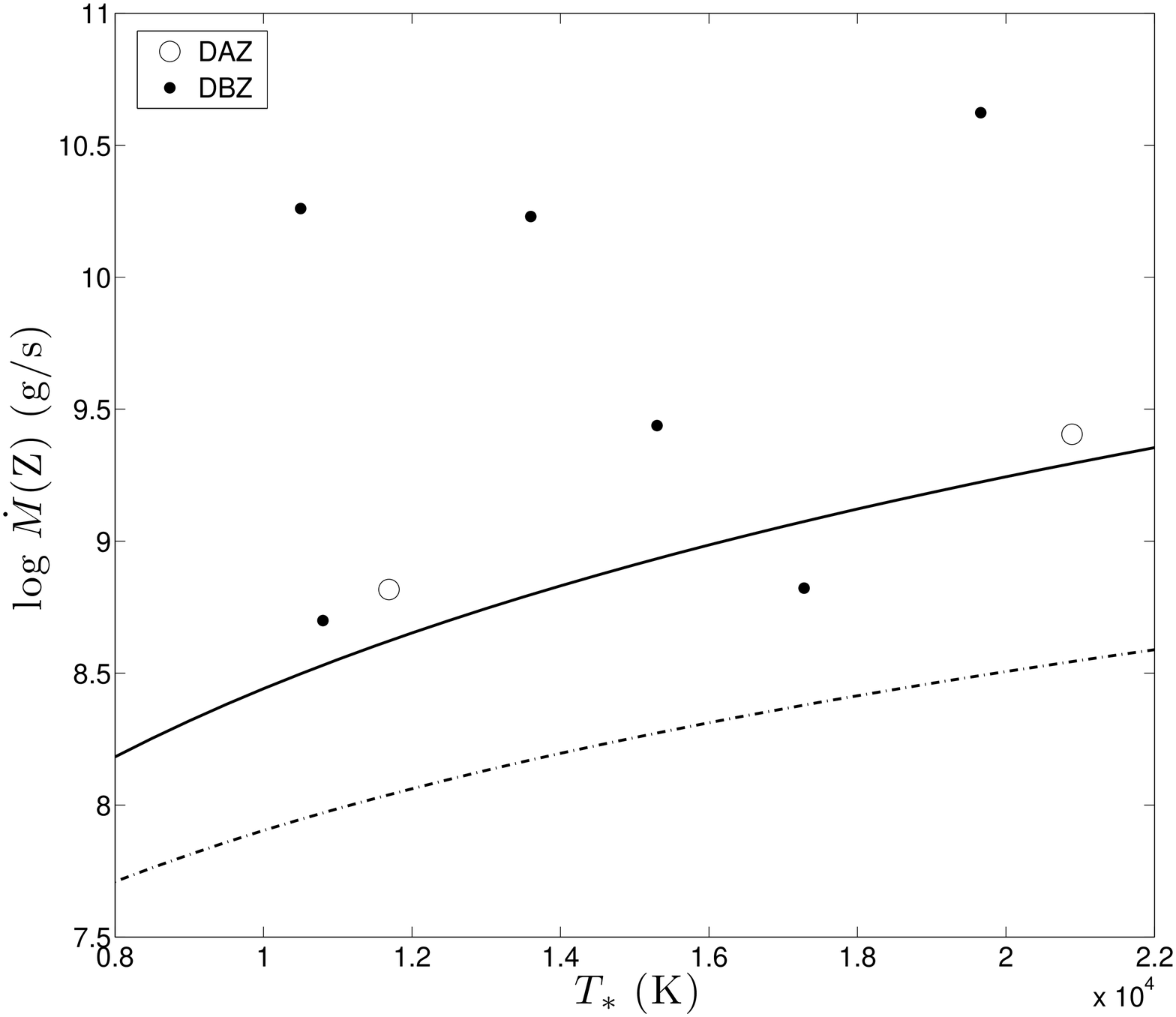}
\caption{Mass accretion rate onto white dwarfs with an infrared excess. The solid line shows the value predicted by our model while the dashed line is the value derived in \citet{Rafikov2011a}. The DAZs are G29-38 \citep{Koester2009a} and GALEX 1931 \citep{Vennes2011}. And DBZs are GD 40 \citep{Klein2010}, GD 61 \citep{Farihi2011a}, SDSS J0738 \citep{Dufour2010}, PG 1225-079 \citep{Klein2011}, Ton 345 (D. Koester, private comminucation) and GD 362 \citep{Koester2009a}. All the stars in this figure have well determined major element abundance. \label{Fig: PRDrag}}
\end{figure}

A comparison between these two accretion rates are shown in Figure \ref{Fig: PRDrag} and we see they differ as much as a factor of 5.

Observationally, we can also derive the accretion rate from the atmospheric pollution. Since we are trying to make a comparison between models that vary by a factor of 5, we need accurate abundances. Therefore, we only consider stars that have measurements for at least three of the four major elements in white dwarf pollution \citep{Klein2010}: O, Mg, Si and Fe.  In Figure \ref{Fig: PRDrag}, we see the accretion rates derived from our model and spectroscopical analysis agree well for DAZs. Since the settling time for heavy elements is relatively short in these stars, the presence of a dusty disk indicates the accretion should be in a steady state \citep{Koester2009a}, which satisfies the assumption of the calculation above. For DBZs, the settling time is comparatively longer and we cannot determine if the system is in the build-up phase or steady state. The accretion rate driven by PR drag gives a good lower bound in these stars and there might be other mechanisms on top of that, as discussed in \citet{Rafikov2011b}.

\bibliographystyle{apj}
\bibliography{WD.bib}

\begin{thebibliography}{55}
\expandafter\ifx\csname natexlab\endcsname\relax\def\natexlab#1{#1}\fi

\bibitem[{{All{\`e}gre} {et~al.}(2001){All{\`e}gre}, {Manh{\`e}s}, \&
  {Lewin}}]{Allegre2001}
{All{\`e}gre}, C., {Manh{\`e}s}, G., \& {Lewin}, E. 2001, Earth and Planetary
  Science Letters, 185, 49

\bibitem[{{Bohlin} {et~al.}(2011){Bohlin}, {Gordon}, {Rieke}, {Ardila},
  {Carey}, {Deustua}, {Engelbracht}, {Ferguson}, {Flanagan}, {Kalirai},
  {Meixner}, {Noriega-Crespo}, {Su}, \& {Tremblay}}]{Bohlin2011}
{Bohlin}, R.~C., {Gordon}, K.~D., {Rieke}, G.~H., {Ardila}, D., {Carey}, S.,
  {Deustua}, S., {Engelbracht}, C., {Ferguson}, H.~C., {Flanagan}, K.,
  {Kalirai}, J., {Meixner}, M., {Noriega-Crespo}, A., {Su}, K.~Y.~L., \&
  {Tremblay}, P. 2011, ArXiv e-prints

\bibitem[{{Bonsor} {et~al.}(2011){Bonsor}, {Mustill}, \& {Wyatt}}]{Bonsor2011}
{Bonsor}, A., {Mustill}, A.~J., \& {Wyatt}, M.~C. 2011, \mnras, 594

\bibitem[{{Bonsor} \& {Wyatt}(2010)}]{BonsorWyatt2010}
{Bonsor}, A. \& {Wyatt}, M. 2010, \mnras, 409, 1631

\bibitem[{{Brinkworth} {et~al.}(2009){Brinkworth}, {G{\"a}nsicke}, {Marsh},
  {Hoard}, \& {Tappert}}]{Brinkworth2009}
{Brinkworth}, C.~S., {G{\"a}nsicke}, B.~T., {Marsh}, T.~R., {Hoard}, D.~W., \&
  {Tappert}, C. 2009, \apj, 696, 1402

\bibitem[{{Chu} {et~al.}(2010){Chu}, {Gruendl}, {Bil{\'{\i}}kov{\`a}},
  {Riddle}, \& {Su}}]{Chu2010}
{Chu}, Y., {Gruendl}, R.~A., {Bil{\'{\i}}kov{\`a}}, J., {Riddle}, A., \& {Su},
  K. 2010, ArXiv e-prints

\bibitem[{{Chu} {et~al.}(2011){Chu}, {Su}, {Bilikova}, {Gruendl}, {De Marco},
  {Guerrero}, {Updike}, {Volk}, \& {Rauch}}]{Chu2011}
{Chu}, Y.-H., {Su}, K.~Y.~L., {Bilikova}, J., {Gruendl}, R.~A., {De Marco}, O.,
  {Guerrero}, M.~A., {Updike}, A.~C., {Volk}, K., \& {Rauch}, T. 2011, ArXiv
  e-prints

\bibitem[{{Debes} {et~al.}(2011){Debes}, {Hoard}, {Kilic}, {Wachter},
  {Leisawitz}, {Cohen}, {Kirkpatrick}, \& {Griffith}}]{Debes2011}
{Debes}, J.~H., {Hoard}, D.~W., {Kilic}, M., {Wachter}, S., {Leisawitz}, D.~T.,
  {Cohen}, M., {Kirkpatrick}, J.~D., \& {Griffith}, R.~L. 2011, \apj, 729, 4

\bibitem[{{Debes} \& {Sigurdsson}(2002)}]{DebesSigurdsson2002}
{Debes}, J.~H. \& {Sigurdsson}, S. 2002, \apj, 572, 556

\bibitem[{{Dong} {et~al.}(2010){Dong}, {Wang}, {Lin}, \& {Liu}}]{Dong2010}
{Dong}, R., {Wang}, Y., {Lin}, D.~N.~C., \& {Liu}, X.-W. 2010, \apj, 715, 1036

\bibitem[{{Drahus} {et~al.}(2011){Drahus}, {Jewitt}, {Guilbert-Lepoutre},
  {Waniak}, {Hoge}, {Lis}, {Yoshida}, {Peng}, \& {Sievers}}]{Drahus2011}
{Drahus}, M., {Jewitt}, D., {Guilbert-Lepoutre}, A., {Waniak}, W., {Hoge}, J.,
  {Lis}, D.~C., {Yoshida}, H., {Peng}, R., \& {Sievers}, A. 2011, \apjl, 734,
  L4+

\bibitem[{{Dufour} {et~al.}(2007){Dufour}, {Bergeron}, {Liebert}, {Harris},
  {Knapp}, {Anderson}, {Hall}, {Strauss}, {Collinge}, \&
  {Edwards}}]{Dufour2007}
{Dufour}, P., {Bergeron}, P., {Liebert}, J., {Harris}, H.~C., {Knapp}, G.~R.,
  {Anderson}, S.~F., {Hall}, P.~B., {Strauss}, M.~A., {Collinge}, M.~J., \&
  {Edwards}, M.~C. 2007, \apj, 663, 1291

\bibitem[{{Dufour} {et~al.}(2006){Dufour}, {Bergeron}, {Schmidt}, {Liebert},
  {Harris}, {Knapp}, {Anderson}, \& {Schneider}}]{Dufour2006}
{Dufour}, P., {Bergeron}, P., {Schmidt}, G.~D., {Liebert}, J., {Harris}, H.~C.,
  {Knapp}, G.~R., {Anderson}, S.~F., \& {Schneider}, D.~P. 2006, \apj, 651,
  1112

\bibitem[{{Dufour} {et~al.}(2010){Dufour}, {Kilic}, {Fontaine}, {Bergeron},
  {Lachapelle}, {Kleinman}, \& {Leggett}}]{Dufour2010}
{Dufour}, P., {Kilic}, M., {Fontaine}, G., {Bergeron}, P., {Lachapelle}, F.,
  {Kleinman}, S.~J., \& {Leggett}, S.~K. 2010, \apj, 719, 803

\bibitem[{{Dupuis} {et~al.}(1993){Dupuis}, {Fontaine}, {Pelletier}, \&
  {Wesemael}}]{Dupuis1993a}
{Dupuis}, J., {Fontaine}, G., {Pelletier}, C., \& {Wesemael}, F. 1993, \apjs,
  84, 73

\bibitem[{{Eisenstein} {et~al.}(2006){Eisenstein}, {Liebert}, {Harris},
  {Kleinman}, {Nitta}, {Silvestri}, {Anderson}, {Barentine}, {Brewington},
  {Brinkmann}, {Harvanek}, {Krzesi{\'n}ski}, {Neilsen}, {Long}, {Schneider}, \&
  {Snedden}}]{Eisenstein2006}
{Eisenstein}, D.~J., {Liebert}, J., {Harris}, H.~C., {Kleinman}, S.~J.,
  {Nitta}, A., {Silvestri}, N., {Anderson}, S.~A., {Barentine}, J.~C.,
  {Brewington}, H.~J., {Brinkmann}, J., {Harvanek}, M., {Krzesi{\'n}ski}, J.,
  {Neilsen}, Jr., E.~H., {Long}, D., {Schneider}, D.~P., \& {Snedden}, S.~A.
  2006, \apjs, 167, 40

\bibitem[{{Farihi} {et~al.}(2010{\natexlab{a}}){Farihi}, {Barstow}, {Redfield},
  {Dufour}, \& {Hambly}}]{Farihi2010a}
{Farihi}, J., {Barstow}, M.~A., {Redfield}, S., {Dufour}, P., \& {Hambly},
  N.~C. 2010{\natexlab{a}}, \mnras, 404, 2123

\bibitem[{{Farihi} {et~al.}(2008{\natexlab{a}}){Farihi}, {Becklin}, \&
  {Zuckerman}}]{Farihi2008a}
{Farihi}, J., {Becklin}, E.~E., \& {Zuckerman}, B. 2008{\natexlab{a}}, \apj,
  681, 1470

\bibitem[{{Farihi} {et~al.}(2011){Farihi}, {Brinkworth}, {G{\"a}nsicke},
  {Marsh}, {Girven}, {Hoard}, {Klein}, \& {Koester}}]{Farihi2011a}
{Farihi}, J., {Brinkworth}, C.~S., {G{\"a}nsicke}, B.~T., {Marsh}, T.~R.,
  {Girven}, J., {Hoard}, D.~W., {Klein}, B., \& {Koester}, D. 2011, \apjl, 728,
  L8

\bibitem[{{Farihi} {et~al.}(2010{\natexlab{b}}){Farihi}, {Jura}, {Lee}, \&
  {Zuckerman}}]{Farihi2010b}
{Farihi}, J., {Jura}, M., {Lee}, J., \& {Zuckerman}, B. 2010{\natexlab{b}},
  \apj, 714, 1386

\bibitem[{{Farihi} {et~al.}(2009){Farihi}, {Jura}, \& {Zuckerman}}]{Farihi2009}
{Farihi}, J., {Jura}, M., \& {Zuckerman}, B. 2009, \apj, 694, 805

\bibitem[{{Farihi} {et~al.}(2008{\natexlab{b}}){Farihi}, {Zuckerman}, \&
  {Becklin}}]{Farihi2008b}
{Farihi}, J., {Zuckerman}, B., \& {Becklin}, E.~E. 2008{\natexlab{b}}, \apj,
  674, 431

\bibitem[{{Fazio et al.}(2004)}]{Fazio2004}
{Fazio et al.} 2004, \apjs, 154, 10

\bibitem[{{Girven} {et~al.}(2011){Girven}, {Gaensicke}, {Steeghs}, \&
  {Koester}}]{Girven2011}
{Girven}, J., {Gaensicke}, B.~T., {Steeghs}, D., \& {Koester}, D. 2011, ArXiv
  e-prints

\bibitem[{{Hoard} {et~al.}(2007){Hoard}, {Wachter}, {Sturch}, {Widhalm},
  {Weiler}, {Pretorius}, {Wellhouse}, \& {Gibiansky}}]{Hoard2007}
{Hoard}, D.~W., {Wachter}, S., {Sturch}, L.~K., {Widhalm}, A.~M., {Weiler},
  K.~P., {Pretorius}, M.~L., {Wellhouse}, J.~W., \& {Gibiansky}, M. 2007, \aj,
  134, 26

\bibitem[{{Holberg} \& {Bergeron}(2006)}]{HolbergBergeron2006}
{Holberg}, J.~B. \& {Bergeron}, P. 2006, \aj, 132, 1221

\bibitem[{{Jura}(2003)}]{Jura2003}
{Jura}, M. 2003, \apjl, 584, L91

\bibitem[{{Jura}(2008)}]{Jura2008}
---. 2008, \aj, 135, 1785

\bibitem[{{Jura} {et~al.}(2007){Jura}, {Farihi}, \& {Zuckerman}}]{Jura2007b}
{Jura}, M., {Farihi}, J., \& {Zuckerman}, B. 2007, \apj, 663, 1285

\bibitem[{{Jura} \& {Xu}(2012)}]{JuraXu2012}
{Jura}, M. \& {Xu}, S. 2012, AJ, submitted

\bibitem[{{Kilic} {et~al.}(2008){Kilic}, {Farihi}, {Nitta}, \&
  {Leggett}}]{Kilic2008}
{Kilic}, M., {Farihi}, J., {Nitta}, A., \& {Leggett}, S.~K. 2008, \aj, 136, 111

\bibitem[{{Kilic} {et~al.}(2009){Kilic}, {Kowalski}, {Reach}, \& {von
  Hippel}}]{Kilic2009a}
{Kilic}, M., {Kowalski}, P.~M., {Reach}, W.~T., \& {von Hippel}, T. 2009, \apj,
  696, 2094

\bibitem[{{Kilic} {et~al.}(2006){Kilic}, {Munn}, {Harris}, {Liebert}, {von
  Hippel}, {Williams}, {Metcalfe}, {Winget}, \& {Levine}}]{Kilic2006}
{Kilic}, M., {Munn}, J.~A., {Harris}, H.~C., {Liebert}, J., {von Hippel}, T.,
  {Williams}, K.~A., {Metcalfe}, T.~S., {Winget}, D.~E., \& {Levine}, S.~E.
  2006, \aj, 131, 582

\bibitem[{{Klein} {et~al.}(2011){Klein}, {Jura}, {Koester}, \&
  {Zuckerman}}]{Klein2011}
{Klein}, B., {Jura}, M., {Koester}, D., \& {Zuckerman}, B. 2011, \apj, in press

\bibitem[{{Klein} {et~al.}(2010){Klein}, {Jura}, {Koester}, {Zuckerman}, \&
  {Melis}}]{Klein2010}
{Klein}, B., {Jura}, M., {Koester}, D., {Zuckerman}, B., \& {Melis}, C. 2010,
  \apj, 709, 950

\bibitem[{{Koester}(2009)}]{Koester2009a}
{Koester}, D. 2009, \aap, 498, 517

\bibitem[{{Koester} {et~al.}(2011){Koester}, {Girven}, {G{\"a}nsicke}, \&
  {Dufour}}]{Koester2011}
{Koester}, D., {Girven}, J., {G{\"a}nsicke}, B.~T., \& {Dufour}, P. 2011, \aap,
  530, A114+

\bibitem[{{Koester} {et~al.}(2009){Koester}, {Voss}, {Napiwotzki},
  {Christlieb}, {Homeier}, {Lisker}, {Reimers}, \& {Heber}}]{Koester2009b}
{Koester}, D., {Voss}, B., {Napiwotzki}, R., {Christlieb}, N., {Homeier}, D.,
  {Lisker}, T., {Reimers}, D., \& {Heber}, U. 2009, \aap, 505, 441

\bibitem[{{Liebert} {et~al.}(2005){Liebert}, {Bergeron}, \&
  {Holberg}}]{Liebert2005}
{Liebert}, J., {Bergeron}, P., \& {Holberg}, J.~B. 2005, \apjs, 156, 47

\bibitem[{{Martin} {et~al.}(2005){Martin}, {Fanson}, {Schiminovich},
  {Morrissey}, {Friedman}, {Barlow}, {Conrow}, {Grange}, {Jelinsky},
  {Milliard}, {Siegmund}, {Bianchi}, {Byun}, {Donas}, {Forster}, {Heckman},
  {Lee}, {Madore}, {Malina}, {Neff}, {Rich}, {Small}, {Surber}, {Szalay},
  {Welsh}, \& {Wyder}}]{Martin2005}
{Martin}, D.~C., {Fanson}, J., {Schiminovich}, D., {Morrissey}, P., {Friedman},
  P.~G., {Barlow}, T.~A., {Conrow}, T., {Grange}, R., {Jelinsky}, P.~N.,
  {Milliard}, B., {Siegmund}, O.~H.~W., {Bianchi}, L., {Byun}, Y., {Donas}, J.,
  {Forster}, K., {Heckman}, T.~M., {Lee}, Y., {Madore}, B.~F., {Malina}, R.~F.,
  {Neff}, S.~G., {Rich}, R.~M., {Small}, T., {Surber}, F., {Szalay}, A.~S.,
  {Welsh}, B., \& {Wyder}, T.~K. 2005, \apjl, 619, L1

\bibitem[{{Melis} {et~al.}(2011){Melis}, {Farihi}, {Dufour}, {Zuckerman},
  {Burgasser}, {Bergeron}, {Bochanski}, \& {Simcoe}}]{Melis2011}
{Melis}, C., {Farihi}, J., {Dufour}, P., {Zuckerman}, B., {Burgasser}, A.~J.,
  {Bergeron}, P., {Bochanski}, J., \& {Simcoe}, R. 2011, \apj, 732, 90

\bibitem[{{Melis} {et~al.}(2010){Melis}, {Jura}, {Albert}, {Klein}, \&
  {Zuckerman}}]{Melis2010}
{Melis}, C., {Jura}, M., {Albert}, L., {Klein}, B., \& {Zuckerman}, B. 2010,
  \apj, 722, 1078

\bibitem[{{Nesvorn{\'y}} {et~al.}(2010){Nesvorn{\'y}}, {Jenniskens}, {Levison},
  {Bottke}, {Vokrouhlick{\'y}}, \& {Gounelle}}]{Nesvorny2010}
{Nesvorn{\'y}}, D., {Jenniskens}, P., {Levison}, H.~F., {Bottke}, W.~F.,
  {Vokrouhlick{\'y}}, D., \& {Gounelle}, M. 2010, \apj, 713, 816

\bibitem[{{Rafikov}(2011{\natexlab{a}})}]{Rafikov2011a}
{Rafikov}, R.~R. 2011{\natexlab{a}}, \apjl, 732, L3+

\bibitem[{{Rafikov}(2011{\natexlab{b}})}]{Rafikov2011b}
---. 2011{\natexlab{b}}, \mnras, L287+

\bibitem[{{Reach} {et~al.}(2005){Reach}, {Megeath}, {Cohen}, {Hora}, {Carey},
  {Surace}, {Willner}, {Barmby}, {Wilson}, {Glaccum}, {Lowrance}, {Marengo}, \&
  {Fazio}}]{Reach2005b}
{Reach}, W.~T., {Megeath}, S.~T., {Cohen}, M., {Hora}, J., {Carey}, S.,
  {Surace}, J., {Willner}, S.~P., {Barmby}, P., {Wilson}, G., {Glaccum}, W.,
  {Lowrance}, P., {Marengo}, M., \& {Fazio}, G.~G. 2005, \pasp, 117, 978

\bibitem[{{Reid} {et~al.}(2001){Reid}, {Liebert}, \& {Schmidt}}]{Reid2001}
{Reid}, I.~N., {Liebert}, J., \& {Schmidt}, G.~D. 2001, \apjl, 550, L61

\bibitem[{{Steele} {et~al.}(2011){Steele}, {Burleigh}, {Dobbie}, {Jameson},
  {Barstow}, \& {Satterthwaite}}]{Steele2011}
{Steele}, P.~R., {Burleigh}, M.~R., {Dobbie}, P.~D., {Jameson}, R.~F.,
  {Barstow}, M.~A., \& {Satterthwaite}, R.~P. 2011, ArXiv e-prints

\bibitem[{{Vennes} {et~al.}(2011){Vennes}, {Kawka}, \&
  {N{\'e}meth}}]{Vennes2011}
{Vennes}, S., {Kawka}, A., \& {N{\'e}meth}, P. 2011, \mnras, 413, 2545

\bibitem[{{von Hippel} {et~al.}(2007){von Hippel}, {Kuchner}, {Kilic},
  {Mullally}, \& {Reach}}]{VonHippel2007}
{von Hippel}, T., {Kuchner}, M.~J., {Kilic}, M., {Mullally}, F., \& {Reach},
  W.~T. 2007, \apj, 662, 544

\bibitem[{{Werner et al.}(2004)}]{Werner2004}
{Werner et al.} 2004, \apjs, 154, 1

\bibitem[{{Zacharias} {et~al.}(2005){Zacharias}, {Monet}, {Levine}, {Urban},
  {Gaume}, \& {Wycoff}}]{Zacharias2005}
{Zacharias}, N., {Monet}, D.~G., {Levine}, S.~E., {Urban}, S.~E., {Gaume}, R.,
  \& {Wycoff}, G.~L. 2005, VizieR Online Data Catalog, 1297, 0

\bibitem[{{Zuckerman} \& {Becklin}(1987)}]{ZuckermanBecklin1987}
{Zuckerman}, B. \& {Becklin}, E.~E. 1987, \nat, 330, 138

\bibitem[{{Zuckerman} {et~al.}(2003){Zuckerman}, {Koester}, {Reid}, \&
  {H{\"u}nsch}}]{Zuckerman2003}
{Zuckerman}, B., {Koester}, D., {Reid}, I.~N., \& {H{\"u}nsch}, M. 2003, \apj,
  596, 477

\bibitem[{{Zuckerman} {et~al.}(2010){Zuckerman}, {Melis}, {Klein}, {Koester},
  \& {Jura}}]{Zuckerman2010}
{Zuckerman}, B., {Melis}, C., {Klein}, B., {Koester}, D., \& {Jura}, M. 2010,
  \apj, 722, 725

\end{thebibliography}

\begin{deluxetable}{llllllllll}
\tablewidth{0pt}
\tablecaption{Warm Dust Disks and Candidates Discovered Since 2010 \label{Tab: NewWD}}
\tablehead{
\colhead{WD}  & \colhead{Name}  & \colhead{SpT} & \colhead{T$_*$}  & \colhead{V}     & \colhead{Discovery}    & \colhead{Discovery}  &\colhead{Ref}  \nl
& & &\colhead{(K)} & \colhead{(mag)} & \colhead{Year} & \colhead{Telescope} & 
}
\startdata
0106-328	&HE 0106-3253		&DAZ	&15,700		&15.50		&	2010		&	{\it Spitzer}	&	1\\
0307+077	&HS 0307+0746		&DAZ	&10,200		&16.40		&	2010		&	{\it Spitzer}	&	1\\
0435+410	&GD 61				&DBZ	&17,280		&14.80		&	2011		&	{\it Spitzer}	&	2\\
		&J0738+1835	&DBZ	&13,600		&17.82\tablenotemark{d}		&	2010		&	Germini/NIRI	& 3\\
0843+516	&PG 0843+517			&DA		&23,900		&16.15		&	2011		&	{\it Spitzer}	&	4\\
1225-079\tablenotemark{a} &PG 1225-079			&DZAB	&10,500		&14.80		&	2010		&	{\it Spitzer}	&	1\\
		&J2209+1223	&DBZ	&17,300		&17.43\tablenotemark{d}		&	2011		&	{\it Spitzer}	&	4\\
		&GALEX J1931+0117	&DAZ	&20,890\tablenotemark{b}	&14.20		&	2011		&	WISE	&	5\\
2221-165	&HE 2221-1630		&DAZ	&10,100		&16.10		&	2010		&	{\it Spitzer}	&	1\\
\\
\hline
\\
				&SDSS 0753+2447\tablenotemark{c}	&	DA	& 13,400		&	19.21\tablenotemark{d}	&	2011	& UKIRT	& 6, 7\\
		&SDSS 0959-0200\tablenotemark{c}		&	DA	& 12,000		&	18.34\tablenotemark{d}	& 2011	&UKIRT	& 6\\
		&SDSS 1221+1245\tablenotemark{c}	&	DA	& 12,000		&	18.39\tablenotemark{d}	& 2011	&UKIRT	& 6\\
1318+005	&J1320+0018\tablenotemark{c}&DA		&19,600		& 17.30		&	2011		&	UKIRT		& 7	\\
		&J1557+0916\tablenotemark{c}&DA		&22,000		& 18.17\tablenotemark{d}		&	2011		& 	UKIRT		& 7	\\
\enddata
\tablenotetext{a}{ WD 1225-079 is reported in \citet{Farihi2010b} to have a 4$\sigma$ excess in the IRAC 7.9 $\mu$m band.}
\tablenotetext{b}{\citet{Melis2011} derived a stellar temperature of 23,470 K for this star.}
\tablenotetext{c}{ These are debris disk candidates that needs to be confirmed with more infrared data as the UKIRT observation only extends to the K band.}
\tablenotetext{d}{This is SDSS r magnitude.}
\tablecomments{This table is supplementary to Table 1 in \citet{Farihi2009}, which lists 14 white dwarfs with an infrared excess discovered at that time.}
\tablecomments{GD 303, which might have a marginal excess at 2$\sigma$ level at the IRAC 4 band \citep{Farihi2010b}, is excluded from this table.}
\tablerefs{
 (1) \citet{Farihi2010b}; (2) \citet{Farihi2011a}; (3) \citet{Dufour2010}; (4) This work; (5) \citet{Debes2011}; (6) \citet{Girven2011}. (7) \citet{Steele2011}; }
\end{deluxetable}

\begin{deluxetable}{lcclllllll}
\tablewidth{0pt}
\tablecaption{Cycle 7 White Dwarf Targets \label{Tab: Target}}
\tablehead{
\colhead{}  & \colhead{}  & \colhead{T$_*$} & \colhead{SDSS z}  & \colhead{}      & \colhead{}   \nl
\colhead{WD}  &\colhead{Name}  &\colhead{(K)} & \colhead{(mag)} & \colhead{[Ca]/[He]\tablenotemark{a}} & \colhead{Ref} & 
}
\startdata
0033-114	&	J003601.37-111213.8	& 7,280	& 17.18		&-9.26 		& 1\\
0216-095	&	J021836.69-091944.8	& 9,560	& 17.84		&-10.63	  & 1\\
0936+560	&	J093942.29+555048.7	& 8,680\tablenotemark{b}	& 17.18		&-8.51	 & 1\\
		&	J095119.85+403322.4	& 8,370	& 17.75		&-10.27	& 1\\
1035-003	&	J103809.19-003622.5	& 8,200\tablenotemark{c}	& 17.30		&-7.70 & 2\\
1212-023	&	J121456.39-023402.8	& 6,000	& 17.50		&\nodata		& 3\\
1244+498	&	J124703.28+493423.6	& 16,800	& 17.21		&\nodata		& 4\\
		&	J125752.77+425255.1	&16,800	& 17.94		&\nodata			& 4\\
		&	J130905.26+491359.7	& 8,620	& 17.67			&-10.16& 1\\
 		&	G 199-63				&8,900	& 17.15		&-9.39		& 1\\
		&	LP 219-80				&6,770	& 16.91			&-11.35	 & 1\\
		&	J220934.84+122336.5	&17,300	& 17.91		&-6.16\tablenotemark{d}		& 4\\
		&	J222802.05+120733.3	&6,760	& 16.36		&-9.96		& 1\\
2222+683	&	G 241-6				&15,300	& 16.07		&-7.25		& 5\\
\enddata
\tablenotetext{a}{ Abundances are expressed as [X]/[Y] = log [n(X)/n(Y)]; n(X) is the number abundance.}
\tablenotetext{b}{This star is reported to be 11,500 K in \citet{Kilic2008}.}
\tablenotetext{c}{This star is reported to be 6,770 K and [Ca]/[He]= -9.44 in \citet{Dufour2007}.}
\tablenotetext{d}{This value is derived from measuring the equivalent width of Ca II 3933.7{\AA} from the SDSS spectra and extrapolate from GD 61, a heavily polluted white dwarf with similar effective temperature and surface gravity.}
\tablerefs{
(1) \citet{Dufour2007}; (2) \citet{Koester2011}; (3) \citet{Reid2001}; (4) \citet{Eisenstein2006}; (5) \citet{Zuckerman2010}.}
\end{deluxetable}

\begin{deluxetable}{llllllll}
\tablewidth{0pt}
\tablecaption{Archived White Dwarf Targets \label{Tab: ArchiveTarget}}
\tablehead{
\colhead{}  & \colhead{T$_{\rm eff}$} & \colhead{SDSS z}    & \colhead{}   \nl
\colhead{WD}    &\colhead{(K)} & \colhead{(mag)} & \colhead{Ref} & 
}
\startdata
0816+297&	16,700	&16.53	& 1\\
0819+363&	18,700	&16.38	& 1\\
0843+516	&	23,900	&16.86	& 1\\
0937+505&	35,900	&16.86	& 1\\
1017+125&	21,400	&16.55	& 2\\
1109+244&	37,800	&16.64	& 3\\
1120+439&	27,200	&16.28	& 3\\
1133+293&	23,000	&15.69	& 3\\
1214+267&	65,700	&16.54	& 3\\
1216+036&	14,400	&16.66	& 2\\
1257+032&	17,600	&16.39	& 2\\
1507+021&	20,200	&17.23	& 2\\
1553+353&	25,600	&15.59	& 3\\
1559+128&	29,200	&17.74	& 3\\
1620+513&	20,900	&16.61	& 1\\
2120+054&	36,200	&16.93	& 1\\
\enddata
\tablerefs{(1) \citet{HolbergBergeron2006}; (2) \citet{Koester2009b}; (3) \citet{Liebert2005}; }
\end{deluxetable}

\begin{deluxetable}{lcccc}
\tablewidth{0pt}
\tablecaption{IRAC Fluxes for Cycle 7 Targets \label{Tab: Flux}}
\tablehead{
\colhead{}  	& \colhead{F$_{3.6\mu m}$}  & \colhead{F$_{4.5\mu m}$}  \nl
\colhead{Name}  &\colhead{$\mu$Jy}  &\colhead{$\mu$Jy} 
}
\startdata
WD 0033-114	&	61$\pm$4	& 41$\pm$3 \\
WD 0216-095	&	28$\pm$2 & 17$\pm$2 \\
WD 0936+560	&	60$\pm$4 &  37$\pm$3 \\
J0951+4033	&	34$\pm$3 &  22$\pm$2 \\
WD 1035-003	&	53$\pm$3 &  36$\pm$3 \\
WD 1212-023	&	85$\pm$5 &  60$\pm$4	\\
WD 1244+498	&	38$\pm$3 &  23$\pm$2	\\
J1257+4252	&	19$\pm$2	&  13$\pm$2	\\
J1309+4913	&	38$\pm$3&   24$\pm$2	\\
G 199-63	&	57$\pm$4&   37$\pm$3	\\
LP 219-80 &	104$\pm$7 & 71$\pm$5	\\
J2209+1223	&	88$\pm$5 &  86$\pm$5	\\
J2228+1207	&	156$\pm$9 & 104$\pm$6	\\
WD 2222+683	&	95$\pm$5 &	64$\pm$4	 \\
\enddata
\end{deluxetable}

\begin{deluxetable}{lcccc}
\tablewidth{0pt}
\tablecaption{IRAC Fluxes for Archived DAs\label{Tab: ArchiveFlux}}
\tablehead{
\colhead{}  	& \colhead{F$_{3.6\mu m}$}  & \colhead{F$_{4.5\mu m}$} & \colhead{F$_{5.7\mu m}$}  & \colhead{F$_{7.9\mu m}$} \nl
\colhead{WD}  &\colhead{$\mu$Jy}  &\colhead{$\mu$Jy} & \colhead{$\mu$Jy} & \colhead{$\mu$Jy} 
}
\startdata
0816+297&	76$\pm$5	&	51$\pm$4	&	28\tablenotemark{a}	&	25\tablenotemark{a}\\
0819+363&	97$\pm$5	&	50$\pm$3	&	66$\pm$8			&	55$\pm$10\\
0843+516	&	140$\pm$12 & 134$\pm$9&	98$\pm$10		& 	154$\pm$13\\
0937+505&	43$\pm$3		&	27$\pm$3	&	22$\pm$7	&	27\tablenotemark{a}\\
1017+125&	71$\pm$6		&	47$\pm$6	&	35\tablenotemark{a}	&	24\tablenotemark{a}\\
1109+244&	57$\pm$5		&	36$\pm$4	&	37\tablenotemark{a}	&	29\tablenotemark{a}\\
1120+439&	83$\pm$5		&	47$\pm$4	&	65$\pm$11		&	54\tablenotemark{a}\\
1133+293&	147$\pm$9 	&	90$\pm$6	&	61$\pm$13		&	53$\pm$16\\
1214+267&	54$\pm$4		&	35$\pm$3	&	32\tablenotemark{a}	&	19\tablenotemark{a}\\
1216+036&	74$\pm$4		&	45$\pm$3	&	34\tablenotemark{a}	&	36\tablenotemark{a}\\
1257+032&	85$\pm$5		&	51$\pm$4	&	34$\pm$11		&	33\tablenotemark{a}\\
1507+021&	40$\pm$3		&	25$\pm$3	&	32\tablenotemark{a}	&	26\tablenotemark{a}\\
1553+353&	167$\pm$9 	&	102$\pm$6&	60$\pm$7			&	38$\pm$9	\\
1559+128&	21$\pm$2		&	12$\pm$2	&	38\tablenotemark{a}	&	21\tablenotemark{a}\\
1620+513&	69$\pm$4		&	43$\pm$3	&	23\tablenotemark{a}	&	26\tablenotemark{a}\\
2120+054&	45$\pm$3		&	27$\pm$3	&	39\tablenotemark{a}	&	23\tablenotemark{a}\\
\enddata
\tablenotetext{a}{3$\sigma$ upper limit.}
\tablecomments{The WISE photometry for WD 0819+363 and WD 1559+128 might be somewhat misleading as the IRAC images show a background source within 5{\farcs}0, which cannot be resolved by WISE and its measurement indicates a substantial infrared excess for these two stars. Confusion is also likely to be a problem for GALEX 1931+0117, \citet{Melis2011} found a significantly lower L-band flux with ground based photometry than \citet{Debes2011} did by using WISE.}
\end{deluxetable}

\begin{deluxetable}{lccccccccccc}
\tablewidth{0pt}
\tablecaption{Disk Model Parameters \label{Tab: DiskPara}}
\tablehead{
\colhead{Name}  & \colhead{T$_*$} & \colhead{r$_*$/D\tablenotemark{a}} & \colhead{T$_{\rm inner}$} & \colhead{T$_{\rm outer}$} & \colhead{r$_{\rm inner}$} & \colhead{r$_{\rm outer}$} & \colhead{cos i} \nl
	&	\colhead{(K)}	&\colhead{(10$^{-12}$)}& \colhead{(K)}	&\colhead{(K)}	&\colhead{(R$_*$)}	&\colhead{(R$_*$)} &
}
\startdata
WD 0843+516	& 23,900	& 3.9 & 1,260	& 640	& 30	& 75	& 0.17	\nl
J2209+1223	& 15,000 & 1.6 & 1,340	& 590	&	15&	45	& 0.08	\nl
		&	& 1.6		& 910	& 710	&25 &	35	& 0.29
\enddata
\tablenotetext{a}{This is the ratio of the stellar radius and its distance to the Sun.}
\end{deluxetable}

\begin{deluxetable}{lcccccc}
\tablewidth{0pt}
\tablecaption{Infrared Flux Produced by Accretion from Cometary Dust \label{Tab: Comp}}
\tablehead{
\colhead{Name}  	&  \colhead{D} & \colhead{log $\dot{M}$(Z)} & \colhead{F$_{4.5\mu m}$($\mu$Jy)} &\nl
				& \colhead{(pc)} 	& \colhead{(g/s)}	& \colhead{predicted\tablenotemark{a}}& \colhead{observed}
}
\startdata
WD 0033-114	&	55		&	8.66	&	6,300	&	41$\pm$3\\
WD 0216-095	&	97		&	7.34	&	120	&	17$\pm$2\\
WD 0936+560	&	65		&	9.41	&	25,000	&	38$\pm$2\\
J0951+4033	&	84		&	7.69	&	310	&	22$\pm$2\\
WD 1035-003	&	49		&	9.65	&	77,000	&	37$\pm$2\\
J1309+4913	&	83		&	7.76	&	370	&	24$\pm$2\\
G 199-63		&	68		&	8.53	&	3,100	&	37$\pm$2\\
LP 219-80		&	45		&	6.61	&	160	&	71$\pm$4\\
J2228+1207	&	34		&	7.96	&	3,200	&	105$\pm$6\\
\enddata
\tablenotetext{a}{These values are the predicted total flux, which is the sum of the photospheric flux and the excess due to accretion from the cometary dust.}
\tablecomments{This table contains all the Cycle 7 stars with reported distance \citep{Dufour2006}.}
\end{deluxetable}

\clearpage

\clearpage

\clearpage

\clearpage

\clearpage

\clearpage

\newpage
\epsscale{0.8}

\clearpage

\newpage

\clearpage

\clearpage

\end{document}